\documentclass[prl,aps,twocolumn,showpacs,superscriptaddress]{revtex4-1}
\usepackage{blindtext}
\usepackage{textcomp} % Degree symbol
\usepackage{hyperref}
\usepackage{graphicx}
\usepackage{bm}
\usepackage[bbgreekl]{mathbbol}
\usepackage{amsmath,dsfont}
\usepackage{color}
\usepackage{comment}
\usepackage{dcolumn}

\usepackage{pdfpages}
\usepackage{pgffor} % for loops

\makeatletter
\AtBeginDocument{\let\LS@rot\@undefined}
\makeatother

\def\supplementfilename{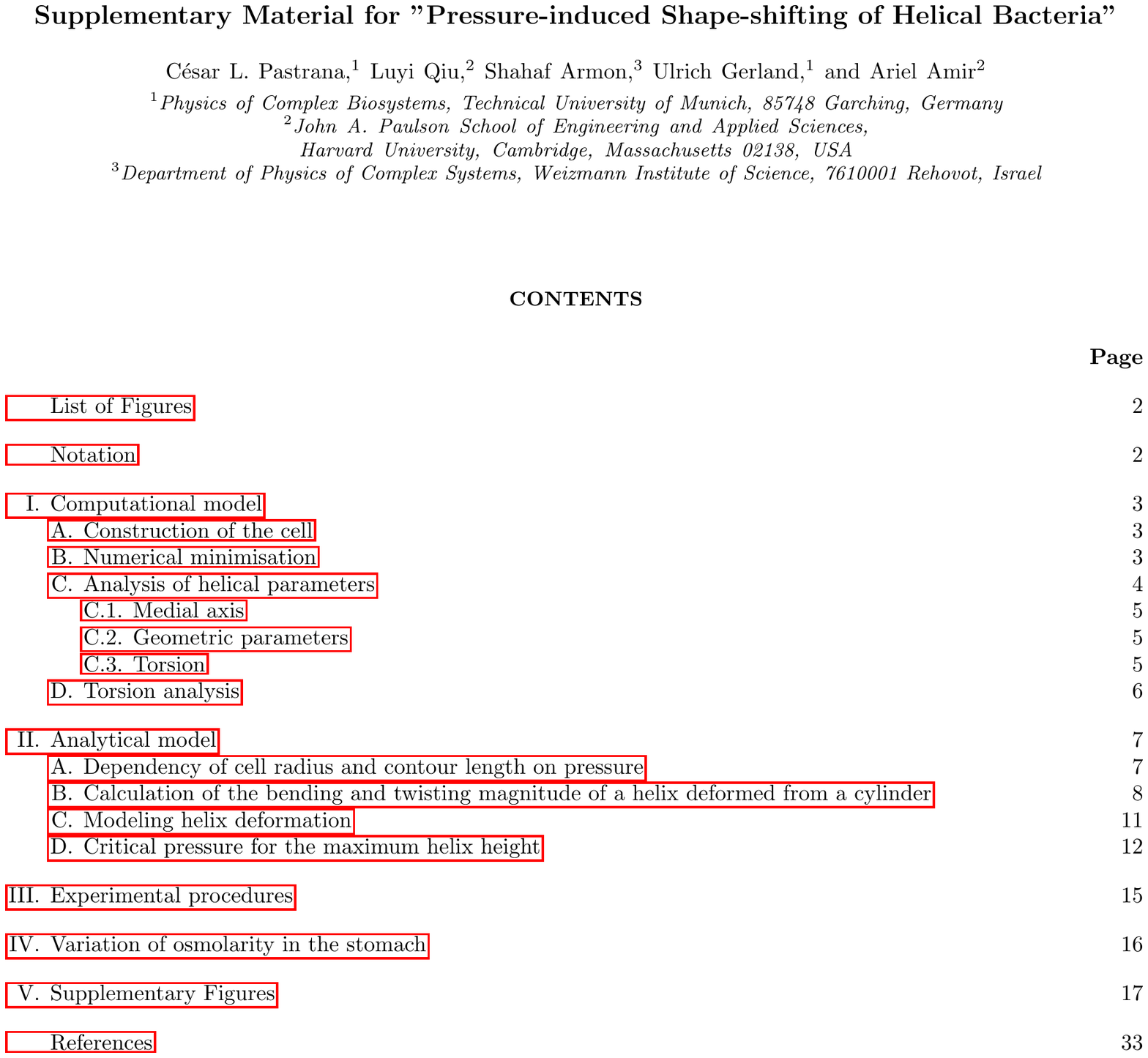}

\pdfximage{\supplementfilename}
\def\numbersupplementpages{\the\pdflastximagepages}

\newif\ifarXiv
\arXivtrue 
\DeclareMathOperator*{\atan}{atan}

\begin{document}

%%%%%%%%%%%%%%% HEADER %%%%%%%%%%
\title{Pressure-induced Shape-shifting of Helical Bacteria}
\author{C\'esar L. Pastrana}
\thanks{These two authors contributed equally}
\affiliation{Physics of Complex Biosystems, Technical University of Munich, 85748 Garching, Germany}
\author{Luyi Qiu}
\thanks{These two authors contributed equally}
\affiliation{John A. Paulson School of Engineering and Applied Sciences, Harvard University, Cambridge, Massachusetts 02138, USA}
\author{Shahaf Armon}
\affiliation{Department of Physics of Complex Systems, Weizmann Institute of Science, 7610001 Rehovot, Israel}
\author{Ulrich Gerland}
\affiliation{Physics of Complex Biosystems, Technical University of Munich, 85748 Garching, Germany}
\author{Ariel Amir}
\affiliation{John A. Paulson School of Engineering and Applied Sciences, Harvard University, Cambridge, Massachusetts 02138, USA}

%%%%%%%%%%%%%%%%%%%%%%%%%%%%%%%%%%

\begin{abstract}
Many bacterial species are helical in form, including the widespread pathogen \emph{H. pylori}. Motivated by recent experiments on \emph{H. pylori} showing that cell wall synthesis is not uniform, we investigate the possible formation of helical cell shape induced by elastic heterogeneity. We show, experimentally and theoretically, that helical morphogenesis can be produced by pressurizing an elastic cylindrical vessel with helical reinforced lines. The properties of the pressurized helix are highly dependent on the initial helical angle of the reinforced region. We find that steep angles result in crooked helices with, surprisingly, reduced end-to-end distance upon pressurization. This work helps to explain the possible mechanisms for the generation of helical cell morphologies and may inspire the design of novel pressure-controlled helical actuators. 
\end{abstract}

\maketitle

%%%%%%%%%%%%%%%%%%%%%%%%%%%%%%%%%%%%%%%%%%%%%%%%%%%    M A I N     D O C U M E N T    %%%%%%%%%%%%%%%%%%%%%%%%%%%%%%%%%%%%%%%%%%%%%%%%%%%%

\section{Introduction}

\begin{figure}[b!]
    \centering
    \includegraphics[width=0.48\textwidth]{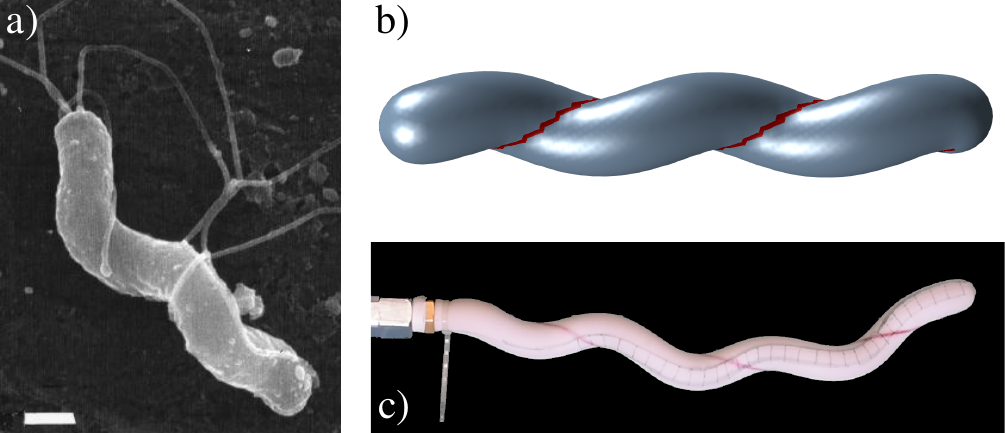}
    \caption{a)  Electron micrograph of the helical bacterium \emph{Helicobacter  pylori}. Scale bar is 0.5 $\mu$m. Image reproduced with permission from \citep{helicobacter_book_physi_gen}. b) Simulated helical cell obtained after pressurizing a spherocylinder, where material is reinforced along a helical path with a reinforcement angle $\alpha = 47.6$\textdegree\hspace{1ex} (corresponding to $n_r$ = 2.5 turns of the helix). The pressure used is such that $\bar{p} := pr_0/Y =$ 0.21 (where $r_0$ is the unpressurized tube radius and $Y$ the 2d Young modulus). c) Silicone balloon reinforced by an (effectively) inextensible thread (red), pressurized to $p \approx$ 0.2 atm.}
    \label{fig:introduction}
\end{figure}

Bacteria span a large range of sizes and a diversity of shapes \citep{selective_value_shape_young}. In most species, cell morphology is primarily determined by the peptidoglycan cell wall, an intricate polymeric mesh that coats the cell and serves as a stress-bearing structure against turgor pressure \citep{koch_growth_book}. Importantly, cells with helical morphology are broadly present across prokaryotic species and are among the most common cellular shapes (Fig.~\ref{fig:introduction}a)\citep{kysela_cladogram_bacteria}. 
In addition to species that naturally grow as helices, mutants of rod-shaped and C-shaped species may acquire helical morphology in septation-deficient mutants \citep{crescentin_curving_growth_cjw}, when septation is suppresed \citep{vibrio_shaevitz}, or under environmental stresses \citep{wortinger_caulobacter_helixes}. These observations might reflect common underlying mechanisms that the cell can adjust to generate different morphologies, allowing the shape-shifting between helix, curved and rod cells.
Conversely, the ubiquity of helical cell shapes in bacteria may indicate a convergent selection of the helical design to serve specific biological functions. For instance, \emph{Helicobacter pylori}, a gastrointestinal pathogen  and paradigmatic example of helical cell, shows reduced pathogenicity in mutants lacking helical shape \citep{helicobacter_book_physi_gen,salama_xlink_relaxation_pg}. \par
Since bacterial cells can transition between rod shape or C-shape to helices, it is natural to ask how cell shape can be tuned to accommodate the conditions of the ever-changing environment characteristic of bacterial life forms. Their ability to control cell shape relies on the regulation of the synthesis and mechanical properties of the cell wall. Current models propose a stress-dependent \citep{koch_growth_book,prl_mechanics_growth,helical_cells_stresses_sun} and/or curvature-dependent \citep{Hussain_mreb_alingment_energy, helicobacter_shaevitz_main, salama_xlink_relaxation_pg} remodeling and growth of the cell envelopes. Moreover, the interaction of cell envelopes with scaffold proteins, particularly in the form of stretch-resistance fibers analogous to the cytoskeleton in eukaryotes, has been proposed as cell shape regulation model, such as the pioneer work by Wolgemuth~\citep{wolgemuth_how_to_make}.  These two mechanisms are not mutually exclusive and very likely work in conjunction \citep{bending_bacteria_mreb_shaevitz, mreb_curvature_feedback}. Both processes can, potentially, lead to a complex envelope composed of regions with different rigidities, akin to a composite material \citep{composite_reinforced, is_crescentin_twist_free_sun}. \par
The underlying microscopic organization of the cell envelope can be manifested upon changes of the internal (turgor) pressure\citep{shaevitz_chiral_twist}. For example, in \emph{E. coli} a decrease of the inner turgor pressure mediated by hyper-osmotic shock results, as expected, in a decrease of the cell's length and radius, though the reduction in radius is minimal\citep{rojas_growh_stress_dependent_osmotic}. This mechanical response has been attributed to the circumferential organization of the cell wall. Conversely, in \emph{H. pylori}, a decrease in turgor pressure results in an increase of length (Fig. S6, ESI\dag, \cite{casademunt_junior_thesis, taylor_phdthesis}). This behavior is not observed in previous mechanical models~\citep{wolgemuth_how_to_make}. 
\begin{figure*}[t]
    \centering
    \includegraphics[width=0.98\textwidth]{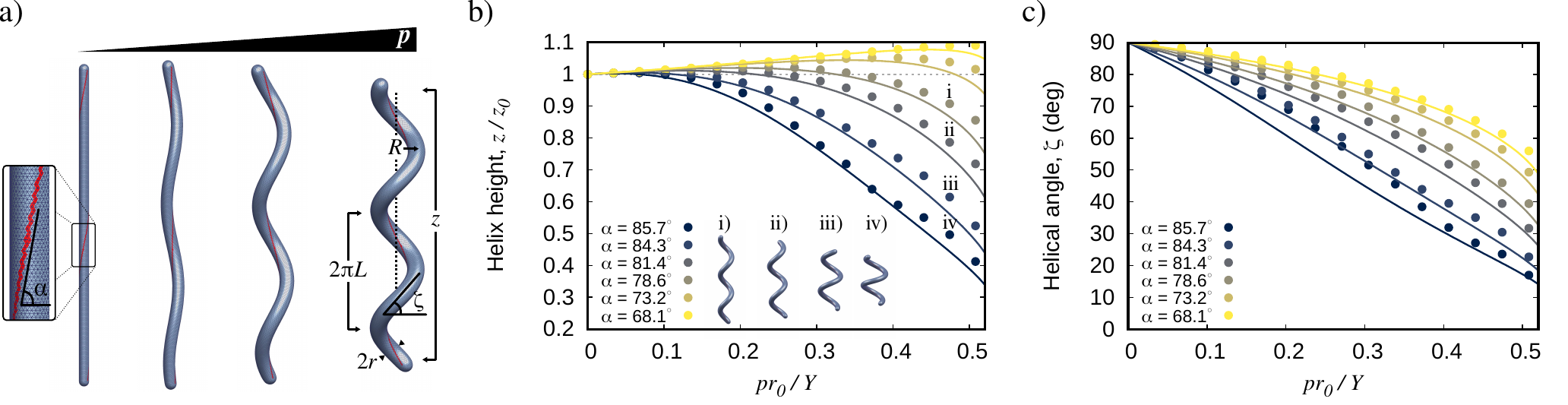}
    \caption{Shape morphology as a function of pressure. a) Spherocylinder with reinforcement along $\alpha =$ 78.6\textdegree\hspace{0.25ex} (corresponding to $n_r=$ 2 turns) at different pressures $p$, from left to right: 0, 0.60, 1.10 and 1.30 atm (where $Y=60$ mN/m and $r_0=200$ nm). b) Helix height $z$ as a function of the nondimensionalized pressure $\bar{p} = pr_0/Y$. Roman numbers: Minimal energy configurations at $\bar{p} =$ 0.47 for the reinforcement angles $\alpha$ = 78.6, 81.4, 84.3 and 85.7 degrees. c) Helical angle $\zeta$ as a function of $\bar{p}$. Symbols are simulation results; solid lines are model predictions.}
    \label{fig:fig2_helical_properties}
\end{figure*}%\twocolumngrid
\par
In this work we explore the changes in shape of helical cells by turgor-pressure-dependent deformation of rod-like cells with a single helical reinforcement (Fig. \ref{fig:introduction}b). We find that the resulting shape after pressurization strongly depends on the helical angle of the reinforced fiber (Fig. \ref{fig:fig2_helical_properties}a). For large helical reinforcement angles, i.e. aligned with the long axis of the rod, pressurization results in the formation of crooked helices with a non-intuitive shortening of the end-to-end distance. We develop a theoretical model that explains the shortening quantitatively and captures qualitatively the resulting helical properties upon pressurization. We confirm our theoretical results experimentally using silicone balloons (Fig. \ref{fig:introduction}c).

\section{Results}
\subsection{Computational model}
To study the resulting shapes of cells subjected to turgor pressure, we construct a triangular spring mesh in the shape of a spherocylinder (Fig.~\ref{fig:fig2_helical_properties}). Under small deformations, this mesh model maps to a continuous material with Poisson's ratio $\nu = \frac{1}{3}$, and 2D Young modulus $Y$ given by $Y=\frac{2}{\sqrt{3}}k$, with $k$ the spring constant of the mesh \citep{seoung_nelson_flexible_membranes}. We assigned a stiffness to the mesh from the characteristic 3D Young modulus $E$ and a thickness $t$ of Gram negative cell walls, where $Y=Et$. Inspired by the experimentally-observed preferential binding of cytoskeleton-like proteins to concave areas of the cell surface \citep{crescentin_curving_growth_cjw, vibrio_shaevitz, helicobacter_shaevitz_main}, we consider a reinforced helical region on the main cylindrical part to be significantly stiffer. This results in a nearly-undeformable region at the range of pressures explored. In real cells, the reinforcement can be a consequence of the anisotropic remodeling of the cell wall, the interaction with protein scaffolds or a combination of both. We find the configurations of minimal energy using a non-linear conjugate gradient algorithm (Sec. I(B), ESI\dag). Upon pressurization, the resulting difference in the resistance to stretching of the reinforced region with respect to the main body leads to the formation of cells with helical morphology  (Fig.~\ref{fig:fig2_helical_properties}a).\par

Naively, we would expect that an increase in the pressure $p$ results in an expansion of the contour length $L_c$ (total length of the tube, caps excluded), the height of the helix $z$ and the radius of the tube $r$ concomitant to the formation of the helix. This is the behaviour observed experimentally in macroscopic models \citep{wolgemuth_how_to_make}. Though the general intuition is correct for $L_c$ and $r$, in contrast, a subtle increase of $z$ during the initial steps of the pressurization is later followed by a shortening, such that $z(p) < z_0$ (Fig. \ref{fig:fig2_helical_properties}a and  \ref{fig:fig2_helical_properties}b). This result is compatible with the increase of the end-to-end distance after a reduction of the turgor pressure, via hyper-osmotic shocking, observed in \emph{H. pylori} (Fig. S6, ESI\dag, \cite{casademunt_junior_thesis, taylor_phdthesis}). We study the shortening behavior in a range of biologically-relevant pressures. We will use a non-dimensional rescaled pressure $\bar{p}:= pr_0/Y$, where $r_0$ is the radius of the non-pressurized tube.\par
The shape after pressurization is highly dependent on the helical angle of the reinforced line, $\alpha$, which is defined in the undeformed configuration.  The shape changes from a tube with helical bumps on the surface to a coiled helix as $\alpha$ increases from 30\textdegree\hspace{0.25em} to 86\textdegree\hspace{0.5em} (Fig. S7, ESI\dag). For $\alpha > 70^\circ$, the resulting shape is well described by a helix and this is the parameter regime we focus on in this letter. We found that for cylinders with a number of turns of the reinforced line $n_r$ greater than 0.5 turns, the resulting helical shape at different pressure is approximately independent of the initial length (Fig. S13, ESI\dag). For the sake of simplicity, we study the formation of helices in the $n_r$ independent regime.\par 

% Description of observed parameters
A finite helix can be defined by four parameters: the radius of the tube $r$, the radius of the helix $R$, the pitch $L$, and the number of turns of the helix central axis $n$. Other variables of interest can be derived from these four parameters, e.g. $L_c = 2\pi n\sqrt{R^2 + L^2}$ (not including the caps). We analyze the response of the helical parameters as a function of pressure (Fig. S8, ESI\dag). We find that $R$ increases when the shell is pressurized to low pressures. However, contrary to $r$ and $L_c$, its dependency with $\bar{p}$ can be non-monotonic. In the large $\alpha$ regime, $R$ shows a well-defined maximum that precedes its decrease when higher pressures are applied. The helical pitch $L$ does not coincide with the pitch of the reinforced line and follows a similar trend to that observed for $z$, more precisely, an increase of less than $15\%$ during the initial steps of the pressurization before a reduction with further increase of $\bar{p}$. A proxy for the shape and crookedness of the helix is given by the helical angle $\zeta = \atan(L/R)$. For every reinforcement angle, $\zeta$ decreases monotonically with $p$, where steep $\alpha$ leads to smaller helical angles for a given pressure (Fig.~\ref{fig:fig2_helical_properties}c). Surprisingly, we note that $n$ is higher than the original number of turns of the reinforced string  $n(p) > n_r$, with a deviation increasing with $\alpha$. This behavior is consistent with the observed absence of coupling between the pitch of the helix and that imposed by the reinforced line. Nonetheless, $n$ shows hardly any variation prior or during the shortening transition. Therefore, initial shortening is mostly dependent on the expansion of $R$. We note however that $n$ shows a significant increase at high pressures. This change of trend is approximately coincident with the the critical pressure at which maximum $R$ is obtained.
\par
%%% FIGURE C-SHAPE->HELIX TRANSITION
\begin{figure}[t]
    \centering
     \includegraphics[width=0.45\textwidth]{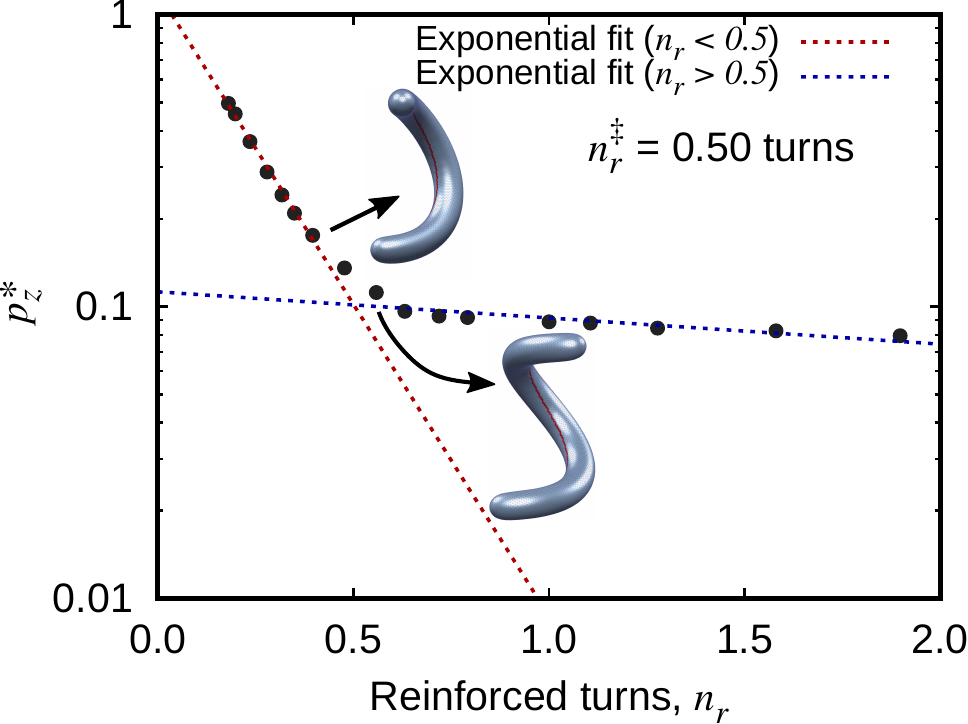}
    \caption{$p_z^*$ as a function of the number of reinforced turns $n_r$. Dashed lines are exponential fits for $n_r < 0.5$ (red) and $n_r > 0.5$ (blue). The intersection between both fits occurs at $n_r^\ddag \simeq 0.5$. \emph{Insets}. Minimal energy configurations for $n_r = 0.40$ and $n_r = 0.59$ showing the soft transition between C-shape to helix ($\bar{p} = $ 0.47). Data for different aspect ratios $L_{c_0}/r_0$, with constant $r_0$, and fixed reinforcement helical angle ($\alpha =$ 84.3\textdegree).}
    \label{fig:c_helix_transition}
\end{figure}
% Torsion
We also examined if surface torsion $\tau$ is involved in the shortening transition (Fig. S9 and Sec. II(B), ESI\dag). We find that $\tau$ grows super-linearly with pressure and that its absolute value for a given pressure is decreasing with $\alpha$. Furthermore, we find that $\tau$ has a smooth dependency on $\bar{p}$ and no change of $\tau$ during pressurization can be associated with a reduction of $z$, thus discarding torsion as a major determinant of the shortening transition. We observe a very similar response of the helical parameters and the torsion upon pressurization in vessels with wider reinforced strings (Fig. S10, ESI\dag).
\par 
We studied the response of shells molded in the shape of helices in the relax configuration. Shells with uniform mechanical properties does not show shortening when pressurized (Fig. S11, ESI\dag). By reinforcing the inner most region of negative Gaussian curvature, we observe a decrease in length and reduction of the helical angle (Fig. S12, ESI\dag).

% C-shape to helix transition
We have shown the pressure-induced formation of helices with steep reinforcement angles $\alpha$. However, the angle $\alpha$ is a function of the number of turns of the reinforced string $n_r$ and the contour length of the tube in its undeformed configuration $L_{c_0}$. We performed simulations with different initial lengths $L_{c_0}$ considering a fixed reinforcement angle ($\alpha = $ 84.3\textdegree), hence varying $n_r$. Only the longer tubes show a length-independent behaviour (Fig. S13). We analyze the response of the critical pressure leading to maximum extension, $p^*_z$, as a function of $n_r$. We observe two clearly distinguishable regimes. For initial lengths that result in $n_r < $ 0.5 we find an exponential decrease of $p^*_z$, whilst for lengths resulting in $n_r > $ 0.5 turns, the critical pressure is mostly insensitive to $n_r$ (Fig. \ref{fig:c_helix_transition}). From visual inspection of the resulting shapes, we note a transition separating a C-shaped-regime for $n_r <$ 0.5 from a helix-dominated regime for $n_r >$ 0.5. Thus, $n_r \approx 0.5$ is a transition point between a C-shaped morphology upon pressurization and a helical shape.

\begin{figure}[b]
    \centering
     \includegraphics[width=0.48\textwidth]{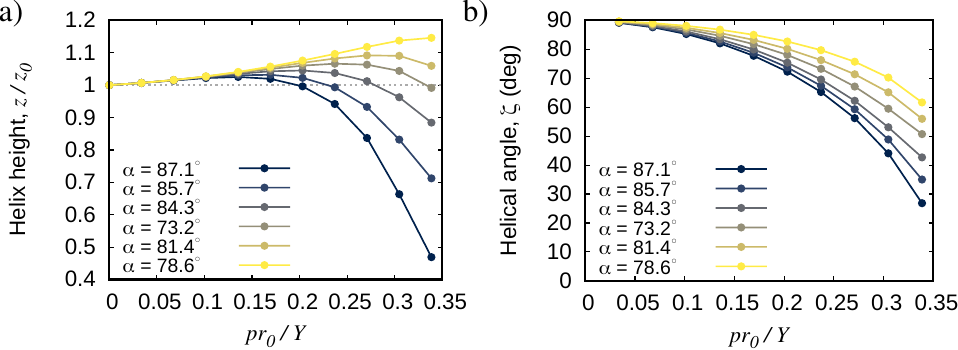}
    \caption{Pressurization of tubes with helical weakening. a) Height of the helix, $z$. b) Helical angle, $\zeta$.  Lines connecting dots are a guide to the eye. The weakening is adjusted by reducing the 2D Young modulus $Y$ of the helical domain by $1/3$. Lines connecting dots are a guide to the eye.}
    \label{fig:weakened_area}
\end{figure}

% Effect of weakening 
The experimental evidence suggest that helical shape may involve the relaxation of the cross-linking bonds of the cell wall \citep{salama_xlink_relaxation_pg, helicobacter_shaevitz_main, predatory_bacteria_relaxation_pg}. Simulations where the helical region is weakened instead of reinforced indicate that weakening is also a feasible mechanism for the formation of helices. Helical parameters during pressurization show a similar response to that observed by reinforcing, including the shortening for large reinforcement angles and the formation of crooked helices (Fig. \ref{fig:weakened_area} and Fig. S14, ESI\dag).

%%% FIGURE WEAKENED AREA
\subsection{Experimental verification}
\begin{figure}[b]
    \centering
     \includegraphics[width=0.45\textwidth]{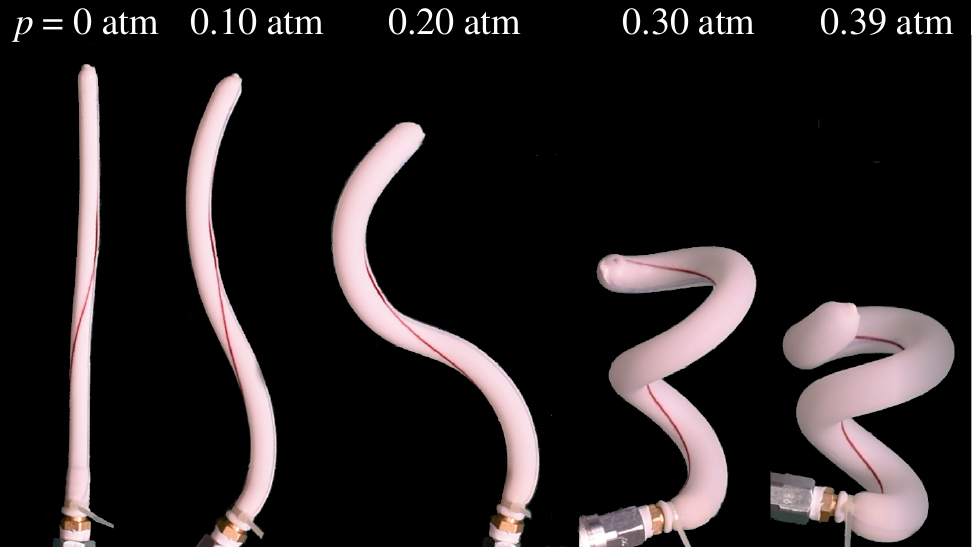}
    \caption{Shape morphology of pressurized silicone models reinforced along a helical path. The string, visible as a red line, is initially set at $\alpha$ = 73\textdegree, and is effectively inextensible. The nondimensionalized pressures used are, from left to right, $\bar{p}$ = 0, 0.38, 0.76, 1.14 and 1.52.}
    \label{fig:balloon_experiments}
\end{figure}

Our numerical model predicts a substantial reduction of $z$ at large reinforcement angles $\alpha$ as a result of the formation of highly crooked helices. To further test this observation, we performed experiments using custom-made silicone balloons with a non-extensible embroidery thread wrapped and embedded on the silicone body (Sec.  III, ESI\dag). We explore pressure-induced formation of helices in the large $\alpha$ regime (Fig~\ref {fig:balloon_experiments}). Silicone shows a hyper-elastic stress-strain response and thus a quantitative comparison with simulation is not possible. Yet, we find qualitative agreement between the experiment and our computational results (Fig. S20, ESI\dag). We observe a similar response of the helical radius $R$ to that observed in the simulations: the increase of $R$ with pressure until reaching a critical point from where subsequent increase in $p$ reduces the helical radius. This is linked to the development of the helix into a more compact structure, resulting in rapid decrease of $z$. Remarkably, the magnitude of the deformations for the indicated helical parameters are on the same scale to that obtained from the numerical simulation.\par 

%%% FIGURE EXPERIMENT

\subsection{Continuum theory}
% Explanation theory
To explain the physics behind the formation of helices, the shortening transition and its dependency on $\alpha$, we model the system analytically by using continuum elasticity. Inspired by the observation that the radius of the tube $r$ and the contour length of the central axis $L_c$ have little dependency on the reinforcement angle (Fig. S15 and Fig. S16, ESI\dag), we assume a separation of the overall swelling of the system and other deformations. Namely, we use an empirical expression for $r$ and $L_c$ as functions of $p$ independent of the reinforcement angle $\alpha$ (Sec.~II(A), ESI\dag). 
The reinforced line always lies along the shortest path with the fixed number of turns on the surface connecting the two ends of the helix to minimize the elastic energy penalty. The analytical expression for this path is provided in Sec.~II(B), ESI\dag~(Equation~S32).
Since the springs in the reinforced region are so stiff that the results do not change with further increase of reinforced spring constant, we assume this reinforced line is unstretchable. This would be equivalent to a constraint that the inner-most part of the helix has a fixed length $L_f$.
We consider the energy per unit length $\Phi$ of a hollow rod subjected to uniform bending and twisting under a linear elasticity framework~\citep{landau_lifshiftz_elasticity}:

\begin{equation}
    \Phi = \frac{EI}{2}\kappa^2 + \frac{\mu J}{2}\tau^2.
    \label{eq:energy_main}
\end{equation}
Here $\tau$ is the twist rate and $\kappa$ is the local curvature of the helix central axis. The parameter $\mu = \frac{E}{2(1+\nu)}$ is the second Lam\'e coefficient, and $I$ and $J$ are the moments of inertia and twist, respectively. For a thin-walled tube, $I = \pi r^3 t$ and $J = 2 \pi r^3 t$, in which $r$ is the radius of the tube and $t$ is the wall thickness. The bending and twisting magnitude of a helix deformed from a cylinder are $\tau = \frac{2\pi n}{L_c}(\sin{\zeta}-n_r/n)$ and  $\kappa = \frac{2\pi n}{L_c}\cos{\zeta}$, see Sec.~II(B) (ESI\dag). 
Again, $\zeta =\arctan{(L/R)}$ is the helical angle, $n_r$ is the number of turns of the reinforced area in the zero pressure case and $n$ is the number of turns of the helix. The state of the system at any pressure can be determined by minimizing the energy given the empirical functions $r(p)$ and $L_c(p)$ under the constraint that the length of the inner-most path of the helix is fixed. This framework allows us to predict, analytically, the helix parameters as a function of pressure (Sec.~II(C), ESI\dag). A comparison between simulation results and the analytical formulation can be found in Fig.~\ref{fig:fig2_helical_properties} and in ESI\dag\hspace{0.5ex}(Fig. S9 and Fig. S17). It is worth noting that in the simulation we observe wrinkles formed near the reinforced string (Fig. S18, ESI\dag). As a consequence of restrained stretching, regions adjacent to the reinforcement string are wrinkled in a direction perpendicular to the string and show well-defined periodicity. The agreement between simulation and theory suggests that energy changes associated with wrinkling are negligible.\par
\begin{figure}[t!]
    \centering
     \includegraphics[width=0.45\textwidth]{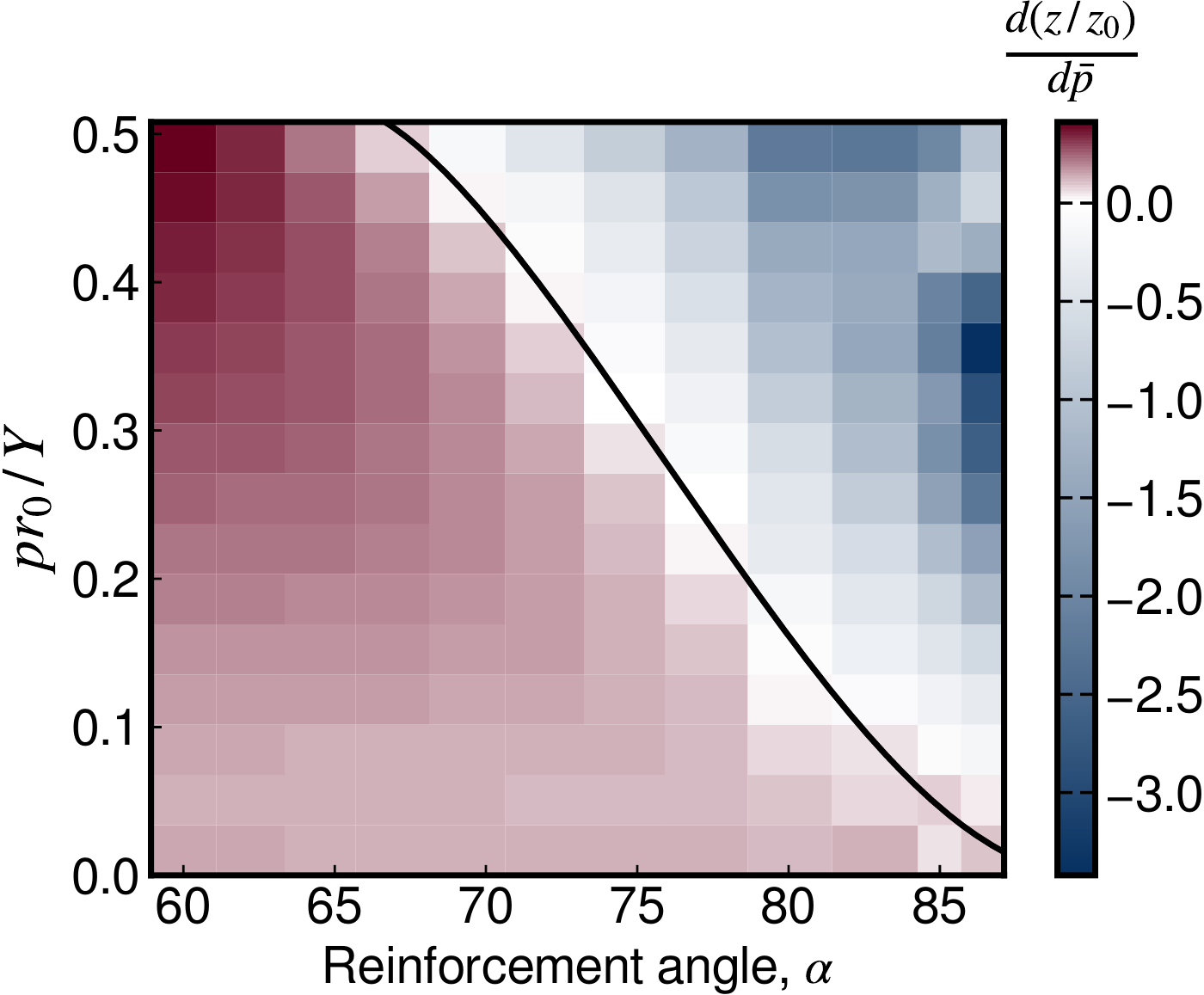}
    \caption{Heatmap of the helix height ($z$) as a function of the angle of reinforcement and the nondimensionalized pressure. The white region corresponds to the region of maximal height observed in the simulation. The black line is the theoretical prediction for the critical pressure at the maximum height $p^*_z(\alpha)$ following Eq. \ref{eq:maximum_height}.}
    \label{fig:phase_diagram_shortening}
\end{figure}

 
%\emph{Shortening transition}.--- 
% Explanation Phase diagram
In Fig.~\ref{fig:phase_diagram_shortening}, we show a phase diagram representing the change in $z$ with respect to the rescaled pressure, as a function of $\alpha$ and $\bar{p}$. The diagram clearly shows the regions of lengthening and shortening in $z$. We can predict analytically the critical pressure of the transition from lengthening to shortening for a given helical angle. 

Due to the constraint that the reinforced path is not stretchable, we have:
\begin{equation}
    L_f =\sqrt{z^2+(2\pi n)^2\cdot (R-r)^2}.
\end{equation}
Since the value of $L_f$ is held fixed when pressure increases, it is then clear that the helix height $z \leq L_f$ and the maximum of $z$ is achieved when $R=r$. 
This corresponds to a configuration where the reinforced line becomes completely straight. 
%When the system starts to get pressurized, $R$ increases from zero as a result of the competition between twisting and bending. If the twisting penalty is zero, e.g. $J$ is set to be zero, $R$ would remain zero to minimize the bending energy. 
%Note that $r$ also increases during pressurization. Thus the state of $R=r$ is not always achievable within the limited range of pressure of biological relevance as we consider here. 
Such straightening behavior of the reinforced string is observed in the simulation: at low pressures, for every $\alpha$, the reinforced string is straightened, increasing its alignment with the long axis of the tube (Fig. S19, ESI\dag). This initial alignment results in the positioning of the reinforced string to the internal (concave) side of the incipient helix.
By requiring $R=r$ we can derive the criteria for the critical pressure at the maximum height under the proper approximations (Sec. II(D), ESI\dag),
\begin{equation}
    \cot^2(\alpha) = \frac{(L_c(p)/L_{c_0})^2-1}{(r(p)/r_0)^2+1},
    \label{eq:maximum_height}
\end{equation}
where $L_{c_0}$ and $r_0$ are the contour length and the tube radius in the undeformed state, and $L_c$ and $r$ are functions of pressure. The right hand side depends on the material properties of the tube, which turns out to be non-linear for the triangular mesh.
Equation \ref{eq:maximum_height} also reveals that the state of maximum height may not be achievable when $\alpha$ is small, as observed in the simulation.
A comparison between Eq.~\ref{eq:maximum_height} and simulation results can be found in Fig.~\ref{fig:phase_diagram_shortening}.

\section{Discussion and Conclusions}
%Conclusions and Discussion.
%%% PHASE DIAGRAM FIGURE
How bacterial shape is generated is an important question in microbiology. Here, we have shown that reinforced rod-shaped cells with the mechanical properties of bacterial cell walls can transit to helical shape upon pressurization. Simulations indicate that large deformations can occur in response to turgor pressure. In particular, large reinforcement angles lead to crooked three-dimensional helices with reduced pitch and large radius, which can result in the shortening of the cellular end-to-end distance. During pressurization, the localization of the reinforced region is reallocated to the internal side of the helix. Then, the experimentally-observed filamentous proteins on the concave region of the cell may reflect strong interactions with the cell wall combined with a resistance to stretching rather than a curvature-dependent binding affinity \cite{crescentin_curving_growth_cjw, non_curvature_dependent_cresc_cjw}.\par
Helical cells display a large variability of the helical parameters, ranging from nearly straight to highly sinuous \citep{helicobacter_shaevitz_main, salama_review_staying_in_shape}. This variability might be attained by precise tuning of the cell wall parameters, namely the helical angle of the reinforced region and the stiffness of reinforced and non-reinforced regions. Interestingly, we find that that helical shapes can be formed by both reinforcement and weakening.

% Discussion
It is generally assumed that a given bacterial species has a characteristic distinct shape. However, bacterial shape is not written in stone and the cell can modify its morphology in response to the environment \citep{salama_review_staying_in_shape}. Changes in morphology can be plastic, by means of cell wall growth, and operating at slow time scales \citep{vibrio_change_shape_QS, salama_xlink_relaxation_pg}. At faster time scales, cell shape can be elastically altered by variations of turgor pressure \citep{shaevitz_chiral_twist, osmolarity_changes_model_pilizota}.
In short, both envelope synthesis (plastic deformation) and elastic deformations contribute to bacterial shape. 
\par

% Length
Our computational results show an example of how cell growth, in combination with elastic deformations, is a potential mechanism to trigger shape transitions. We observe that a number of reinforced turns $n_r \approx 0.5$ turns marks a transition between C-shaped cell and helical cells. Therefore, C-shaped cells ($n_r < $ 0.5 turns) growing in length and keeping a constant $\alpha$ will develop a helical shape as a result of the increase on $n_r$ (Sec. I(E), ESI\dag). This agrees with the experimental observations of C-shaped cells acquiring helical morphology upon inhibition of cell division \citep{vibrio_shaevitz, crescentin_curving_growth_cjw}. Thus, under conditions favouring growth, bacteria can plastically shift from crescent to helical shape \citep{wolgemuth_how_to_make}. 
\par 
Nevertheless, the cell could exploit turgor pressure to elastically and dynamically modify its shape. Human-made soft robots composed of elastomers reinforced with helical fibers, known as McKibben actuators, have been designed to produce a variety of extension, bending and/or twist deformations in response of pneumatic pressures \citep{awn_soft_actuator,bertoldi_robots_match,bertoldi_soft_robots}. Are helical bacteria exploiting turgor pressure to operate as biological McKibben actuators? Variations of turgor pressure could be used to tune cell shape for a specific biological function. In the example of \emph{Helicobacter pylori}, the digestive process leads to variations of the external ionic conditions\citep{beverage_osmolality} (Sec.~IV, ESI\dag) and hence to (transient) variations of turgor pressure. The underlying changes on the helical shape could promote the passive penetration through the mucus of the gut, for instance by means of contraction and expansion combined with variations of the helical radius. The mechanism would resemble that found for the self-burial of the coiled seeds in plants in response to changes of the environmental humidity \citep{erodium_plant_helix}. Alternatively, the cell could use an active mechanism for altering its shape \citep{osmolarity_changes_model_pilizota}: since the ionic conditions of the intracellular medium can be regulated, the cell could, to some extent, alter its osmolarity to adjust cell morphology and the ensuing drilling of the gut. Shifts in turgor pressure in combination with the anisotropic properties of the cell wall have been proposed as drivers of the Venus fly trap snap or the rapid folding of the leaves in \emph{Mimosa pudica} \citep{mahadevan_fly_trap, review_plant_motion}. It is plausible to think that bacteria could use similar mechanisms to that used by plants to drive morphological adaptations and dynamical changes of shape. 
% => Proper term for plants (Nastic motion)
%({\color{red}Cartoon? Important.  Some conflict with stress-stiffening model, see Deng et al})
Our numerical observations, validated with macroscopic experimental models, are provoking for verification in live bacterial cells. Carefully-designed experiments combining fluorescence microscopy with microfluidics devices have permitted to visualize the response of bacteria to changes in turgor pressure \citep{rojas_growh_stress_dependent_osmotic, pilizota_shaevitz_change_shape}. Hence, it is feasible to characterize the instantaneous change of the cell helical properties in response to osmotic shocks.  The cross-fertilization between experiments and simulations will aid in the understanding of the mechanisms driving cell morphology.
 
\section*{Appendix}
In this appendix we list the main parameters and observables used in this work.
\begin{table}[h]
\small
  \caption{Geometric parameters and observables}
  \label{tbl:geometric_paramter}
  \begin{tabular*}{0.48\textwidth}{@{\extracolsep{\fill}}cl}
    \hline
    Parameter & Description\\
    \hline
     $z$ & Helix height\\
     $r$ & Tube radius\\ 
     $L$ & Helical pitch (tube)\\
     $R$ & Helix radius\\
     $\alpha$ & Helical angle (reinforced string)\\
     $\zeta$ & Helical angle (tube)\\
     $L_c$ & Contour length (tube, caps not included)\\
     $L_f$ & Contour length (reinforced string)\\
     $n$ & Number of turns (tube)\\
     $n_r$ & Number of turns (reinforced)\\
     $\tau$ & Torsion of the tube's surface\\
     $\kappa$ & Curvature tube's central axis\\
    \hline
  \end{tabular*}
\end{table}
\begin{table}[h]
\small
  \caption{Mechanical parameters}
  \label{tbl:mechanical_parameters}
  \begin{tabular*}{0.48\textwidth}{@{\extracolsep{\fill}}cl}
    \hline
    Parameter & Description\\
    \hline
     $p$ & Pressure\\
     $E$ & 3D Young's modulus\\
     $Y$ & 2D Young's modulus\\
     $t$ & Shell thickness\\ 
     $I$ & Moment of inertia\\ 
     $J$ & Moment of twist\\
    \hline
  \end{tabular*}
\end{table}

\section*{Acknowledgements}

This work was supported by Volkswagen Stiftung (A.A. and U.G.), NSF CAREER Grant No. 1752024 (A.A.), the NSF-Simons Center for Mathematical and Statistical Analysis of Biology at
Harvard, award number 1764269, the Harvard Quantitative Biology Initiative and Grant NSF-1806818 (L.Q.). The authors acknowledge Joshua W. Shaevitz and Hillel Aharony for useful discussions.

%%%END OF MAIN TEXT%%%

%%%REFERENCES%%%
%%%% BIBLIOGRAPHY
\bibliographystyle{apsrev4-1} % Tell bibtex which bibliography style to use
\bibliography{mainbib} %You need to replace "rsc" on this line with the name of your .bib file
%\includepdf{./Supplementary_Material/supplementary_material}

\ifarXiv
    

\section*{Supplementary material}

# Supplementary Material for "Pressure-induced Shape-shifting of Helical Bacteria"

César L. Pastrana,[1] Luyi Qiu,[2] Shahaf Armon,[3] Ulrich Gerland,[1] and Ariel Amir[2]

[1]*Physics of Complex Biosystems, Technical University of Munich, 85748 Garching, Germany*
[2]*John A. Paulson School of Engineering and Applied Sciences,*
*Harvard University, Cambridge, Massachusetts 02138, USA*
[3]*Department of Physics of Complex Systems, Weizmann Institute of Science, 7610001 Rehovot, Israel*

## CONTENTS

## LIST OF FIGURES

## NOTATION

| Symbol | Parameter |
| --- | --- |
| $p$ | Pressure |
| $p_x^*$ | Pressure at the critical point of $x$ |
| $z$ | Helix height |
| $L_c$ | Contour length (tube's central axis) |
| $L_f$ | Contour length (reinforced string) |
| $r$ | Tube radius |
| $R$ | Helix radius |
| $L$ | Helical pitch (tube) |
| $L_r$ | Helical pitch (reinforced string) |
| $\zeta$ | Helical angle (tube) |
| $\tau$ | Torsion of the tube's surface |
| $\alpha$ | Helical angle (reinforced string) |
| $n$ | Number of turns (tube) |
| $n_r$ | Number of turns reinforced line |
| $t$ | Shell thickness |
| $E$ | 3D Young's modulus |
| $Y$ | 2D Young's modulus |
| $\nu$ | Poisson's ratio |
| $\mu$ | Second Lamé coefficient |
| $I$ | Moment of inertia for thin shell |
| $J$ | Moment of twist for thin shell |

# I. COMPUTATIONAL MODEL

## A. Construction of the cell

The main tube defining the cell is generated by positioning vertices on the surface of a cylinder to obtain a perfectly-ordered equilateral triangular mesh organized in consecutive rings. Two hemispherical caps are added to the top and the bottom of the cylinder matching the radius of the main body. The generation of each cap starts by placing point particles randomly on a hemispherical shell. The number of particles per cap is calculated to have an average distance between particles equal to that of the main body. After random positioning the vertices on the spherical shell, their location is equilibrated by means of a Monte Carlo simulation considering a repulsive interaction between particles given by a Weeks-Chandler-Andersen potential. We obtain the final mesh from the Delaunay-based algorithm Advancing Front Surface Reconstruction routine on the Computational Geometry Algorithms Library (CGAL). We confirmed that the specific topology of the mesh does not affect our results by generating the cylindrical main body following the same protocol used for the caps. This results in vertices located randomly and presenting multiple topological defects, yet organized as nearly-equilateral triangles. We found the same helical properties for both meshes. Helical meshes to be used as input (relax) configuration were generated following the same procedure here described for the random cylinders but considering the parameterization of helix.

The selection of vertices for the helical reinforcement relies on finding the best match between a theoretical helix parameterized as a function of its arc length and the cumulative distance between consecutively-connected vertices. That is, the $i$th vertex of the helix is obtained considering the helix:

$$\mathbf{r}_h = r_0 \cos{(il_0 c)}\,\mathbf{i} + r_0 \sin{(il_0 c)}\,\mathbf{j} + Lil_0 c\mathbf{k},\tag{1}$$

where $r_0$ and $L$ are the radius and the pitch in the undeformed configuration and $c = a(r_0^2 + L^2)^{-1/2}$. The parameter $a = \sqrt{3}/2$ accounts for the mesh geometry and $l_0$ is the bond length of the mesh. The vertex $i$ is selected among the coordinates of the nearest-neighbours of vertex $i - 1$, $\mathbf{r}_i$, as that with the minimum distance to the theoretical helix, $\arg\min_i \|\mathbf{r}_h - \mathbf{r}_i\|$. Concentric vertices to the main helix vertices can be selected for edge reinforcement, increasing the width $w$ of the reinforced helical stripe (Fig. S10a). The results in the main section were obtained considering the minimal width possible given the degree of refinement of our mesh.

The dimensions of the spherocylinder and the mechanical properties of the cell can be found in Table S1.

## B. Numerical minimisation

In our simulation toolbox, the total energy of the pressurized shell is described by:

$$E = E_s + E_b - pV,\tag{2}$$

with $E_s$ the stretching energy, $p$ is the pressure difference and $V$ the volume enclosed by the shell. The stretching energy of the mesh is given by the harmonic potential

$$E_s = \frac{k_s}{2}\sum_{\langle i,j\rangle}(r_{ij} - r_{ij}^0)^2,\tag{3}$$

where the sum runs over all pairs of connected notes $i, j$ constituting an edge of the mesh and $r_{ij} = |\mathbf{r}_i - \mathbf{r}_j|$. Despite the simplicity of this discrete model, it maps to an (equivalent) continuous elastic medium with 2D Young modulus

$$Y = \frac{2}{\sqrt{3}}k_s,\tag{4}$$

and with a constant Poisson ratio imposed by the triangular geometry of the mesh, given by $\nu = 1/3$ [1]. The 2D Young modulus $Y$ is related to the 3D Young modulus $E$ by $Y = Et$, with $t$ the thickness of the shell. The stiffness of the edges connecting vertices identified as part of the helical region is reinforced by a factor $K = 10^3$. This large value was selected to ensure the insensibility of the reinforced region and should not be confused with a requirement for such specific increase in stiffness on a real system.

The bending stiffness is given by

$$E_b = k_b \sum_{\langle \alpha,\beta\rangle}(1 - \hat{\mathbf{n}}_\alpha \cdot \hat{\mathbf{n}}_\beta),\tag{5}$$

| Parameter | Value |
|---|---|
| **Cell configuration** | |
| Cell radius (unpressurized), $r_0$ | 0.20 $\mu m$ |
| Cell length, $L_{c_0}$ | 12.5 $\mu m$ |
| Helix angle, $\alpha$ | 40-88° |
| Number of vertices, $N$ | 10712 |
| **Mechanical parameters** | |
| Cell wall Young modulus (3D), $E$ | 30 MPa |
| Cell wall thickness, $t$ | 2.0 nm |
| Reinforcing factor helical region, $K$ | $10^3$ |
| Turgor pressure, $p$ | 0.1-1.5 atm |

TABLE S1. Simulation parameters. The mechanical parameters were selected to match the described values of Gram negative bacteria [4–6]. The unpressurised radius results in $r$ between 0.20-0.57 $\mu m$ upon pressurisation (0.0-1.0 atm), in the range of experimentally-observed values for *Helicobacter pylori* [7].

where the sum is over the pair of triangles $\alpha, \beta$ sharing an edge, $\hat{\mathbf{n}}$ are their respective normal vectors and $k_b$ is a bending stiffness. For a cylindrical object, the following relation between the latter and the continuous bending rigidity $\kappa$ is [1, 2]:

$$\kappa = \frac{\sqrt{3}}{2} k_b. \tag{6}$$

where the $\kappa$ is a function of the Young modulus and the thickness [3]:

$$\kappa = \frac{Eh^3}{12(1-\nu^2)}. \tag{7}$$

Hence, the discrete stretching and bending stiffnesses of the mesh model are defined by considering the Young modulus and thickness of Gram negative bacteria. Importantly, in Gram negative bacteria the thickness is small compared with radius of the cell. In such case, the stretching energy is much larger than the bending energy, $E_s \gg E_b$ and thus the bending contribution is mostly negligible in non-reinforced regions. We consider that the bending stiffness $k_b$ of a pair of triangles sharing and edge connecting vertices identified as part of the helical region is reinforced by a factor $K$.

To calculate the volume $V$ enclosed by a non-convex polyhedron we use the method described in [8]. The total volume enclosed by the surface is given by

$$V = \frac{1}{6} \sum_{t=N_t} \det(\mathbf{T}_t), \tag{8}$$

where the sum is over all the triangles of the surface $N_t$. The matrix $\mathbf{T}$ is constructed from the coordinates of the vertices defining the triangle, $\mathbf{T} = [\mathbf{v}_1, \mathbf{v}_2, \mathbf{v}_3]$ with $\mathbf{v}_i^T = [x_i, y_i, z_i]^T$ the coordinates of the vertex $i$.

We minimize the total energy (Eq. 2) using a custom non-linear conjugate gradient written in C++ using the secant-method for line search [9]. We consider analytical expressions for all the gradients. The total minimization routine for a final objective pressure is performed by iteratively increasing the pressure in a step-wise fashion until convergence. The code is freely available on the public repository in [10].

The mechanical parameter of the model can be found in Table S1.

### C. Analysis of helical parameters

A three dimensional helix can be described by the radius of the tube $r$, as well as by the pitch $L$, the radius $R$ and the total number of turns $n$ of the helix. Additional quantities such as the contour length of the helix $L_c$, the height $z$ or the end-to-end distance can be inferred from the former parameters. All the aforementioned parameters can be extracted from the medial axis of the helix. Therefore, we relied in extracting the medial axis as the first step in the characterization of the helix geometry.

### C.1. Medial axis

We start by aligning the cell to the $\mathbf{k}$ axis by means of the Rodrigues' rotation formula. To determine the coordinates of the nodes conforming the medial axis $\mathbf{x}_i$ we used a tailored adaptation of the shrinking-ball algorithm [11]. This algorithm is general: it does not depend on any particular organization or topology of the mesh. However, we can exploit the specific topology of the mesh used in our simulations, where the vertices of the tube are organized as consecutive hoops. In this case, we can use the center of mass of each ring as a node of the medial axis. We obtained the same helical parameters by using the shrinking ball algorithm or the center of mass of the hoops for the determination of the medial axis (difference $< 5\%$).

### C.2. Geometric parameters

From the nodes defining the medial axis of the tube, we determine the radius of the tube $r$ as the distance between a vertex of the mesh to i) the line defined by the two nearest medial axis nodes to the given vertex (for the shrinking-ball medial axis) or ii) to the corresponding hoop center of mass (for hoop-based medial axis). Though the shrinking ball algorithm returns $N$ medial axis nodes as well as the radius of the inscribed ball on the tube for each vertex, we noted that for the degree of refinement of our mesh this approach tends to underestimate the real radius of the tube.

The radius of the helix $R$ is given by the radius of a node in the plane perpendicular to the axial direction of the helix $\mathbf{x}_i^\perp$. The local radius of the helix is then:

$$R = ||\mathbf{x}_i^\perp|| \tag{9}$$

with $\mathbf{x}_i^\perp = \mathbf{x}_i - (\mathbf{x}_i \cdot \hat{\mathbf{k}})\hat{\mathbf{k}}$. In a similar manner, we calculate the pitch $L$ as

$$L_i = \frac{z_i}{\theta_i}, \tag{10}$$

where $z_i = \mathbf{t}_i \cdot \hat{\mathbf{k}}$ where $\mathbf{t}_i = \mathbf{x}_i - \mathbf{x}_{i-1}$. The angle $\theta$ represents the rotation on the plane normal to $\hat{\mathbf{k}}$. It is obtained as $\theta_i = \cos^{-1}(\hat{\mathbf{x}}_i^\perp \cdot \hat{\mathbf{x}}_{i-1}^\perp)$. The cumulative angle $\Theta$ is the total number of turns of the helical vessel, $\Theta = \sum \theta_i$ (and then $n = 2\pi\Theta$). The cumulative increase in extension $Z = \sum z_i$ as a function of the cumulative angle is fitted to a linear function to obtain an average pitch. Still, our approach allows to characterize the properties of the helices defined locally.

### C.3. Torsion

In this work we refer as torsion to the twist of the surface of the tube, i.e. the rate of rotation of the material frame along the tangent to the medial axis, and it should not be confused with the torsion of a one dimensional helix [12]. We determine the torsion of the surface by taking advantage of the geometry of the mesh, composed of consecutive hoops organized to form a perfect hexatic structure. Hence, in the undeformed configuration, a vertex at hoop $h$, interacts with two vertexes of the hoop $h + 1$. We call $\mathbf{r}_i$, $\mathbf{r}_i^r$ and $\mathbf{r}_i^l$ to the coordinates of a given vertex and its two upper neighbours to its right and to its left, respectively. We make these points coplanar by subtracting the coordinates of the medial axis at $h$ (for $\mathbf{r}_i$) or at $h+1$ (for $\mathbf{r}_i^l$ and $\mathbf{r}_i^r$). In the undeformed configuration $||\mathbf{r}_i^l - \mathbf{r}_i|| \equiv ||\mathbf{r}_i^r - \mathbf{r}_i|| = l_0/2$, with $l_0$ the bond length of the mesh. Thus, we can determine the torsion by minimizing the objective function for the distance between vertexes:

$$f(\phi) = \sum_{i=1}^{N_h}(||\mathbf{r}_i - \mathcal{R}\mathbf{r}_i^r|| - s)^2 + \sum_{i=1}^{N_h}(||\mathbf{r}_i - \mathcal{R}\mathbf{r}_i^l|| - s)^2, \tag{11}$$

where the sum runs over all the vertices $i$ of a hoop $h$, and the matrix $\mathcal{R} = \mathcal{R}(\phi, \hat{\mathbf{t}}_h)$ is a rotation matrix around the tangent vector to the medial axis $\hat{\mathbf{t}}_h$ at the hoop $h$. The value of the minimum is $s = \pi r_h/N_h$, being $r_h$ the radius at hoop $h$. We used the golden-section search algorithm bracketed on the interval $[-\pi, \pi]$ to determine the argument of the minimum $f(\phi)$. This procedure returns the local torsion between two hoop $h$ and $h + 1$. We find the final torsion of the object from a linear fit to the cumulative sum of the torsion $\Phi = \sum \phi_i$ for all the consecutive hoops as a function of the cumulative length between hoops in the *undeformed* configuration, $\tau = \Phi/L_{c_0}$.

### D.   Torsion analysis

We examined if surface torsion $\tau$ is involved in the relocation of the reinforced string to the internal region of the helix (Fig. S9). We find that $\tau$ grows super-linearly with pressure and its absolute value, for a given pressure, is inversely proportional to the reinforcement angle $\alpha$. This is consistent with the expected limit of no-torsion (only bending) for a reinforced tube with $\alpha = 90°$. Pressurization for low $\alpha$ is torsion-dominated whilst large $\alpha$ is bending-dominated, yet the contribution of the complementary term is not negligible in terms of the total energy. Importantly, simulations show that $\tau$ has a smooth dependency on $\bar{p}$ and that no change of $\tau$ during pressurization can be associated with a reduction of $z$, thus discarding torsion as a major determinant of the shortening transition.

We tested these observations experimentally: we clearly observe torsion of the surface upon pressurization for $\alpha = 50°$ (Fig. S20), in agreement with our computational observations. The separation of a bending dominated and a torsion dominate regime are in agreement with the results obtained by others in fibers-reinforced soft actuators [13, 14].

## II. ANALYTICAL MODEL

### A. Dependency of cell radius and contour length on pressure

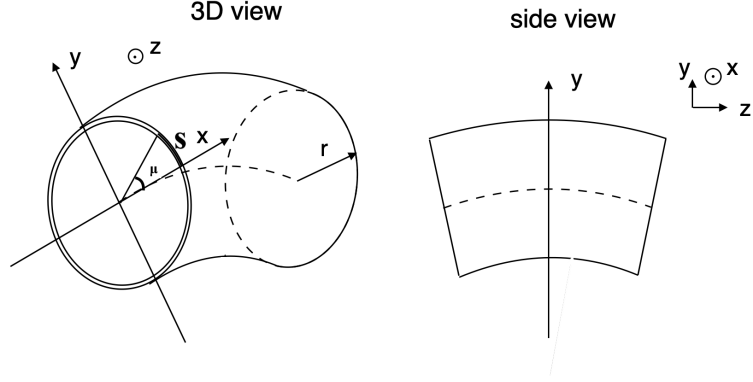

FIG. S1. Parametrization in curved tubes: 3D and side views.

Inspired by the observation from the simulation that the radius of the tube $r$ and the contour length of the central axis $L_c$ have little dependency on the reinforcement angle $\alpha$ (Figure S14), we use an empirical expression for $r(p)$ and $L_c(p)$ in the main text as functions of $p$ (independent of $\alpha$). Here we present a model for the expression of functions $r(p)$ and $L_c(p)$. Since the results hardly depend on the helical reinforcement angle $\alpha$, we begin with an extreme case in which the reinforced line is completely straight, lying in the longitudinal direction of the cylinder. Thus after pressurization the vessel will curve into a banana shape as illustrated in Fig. S1. We assume that the length of this reinforced straight line remains fixed and the curved tube after pressurization has a uniform radius.

We begin with the general elastic energy expression for a continuous material:

$$\Phi = \frac{1}{2} \int N^{\alpha\beta} E_{\alpha\beta} d\xi_1 d\xi_2 \; , \tag{12}$$

with

$$N^{\alpha\beta} = \frac{Et}{1-\nu^2} \left[ (1-\nu) E^{\alpha\beta} + \nu E^\gamma_\gamma g^{\alpha\beta} \right] \; . \tag{13}$$

Here we consider only the stretching energy and neglect the bending energy. We use the parametrization of $\xi_1 = s$ and $\xi_2 = z$ as shown in Fig. S1. $s$ is the distance measured along the curve of the cross-section in the undeformed state and $z$ is the distance along the longitudinal direction of the tube. We further define $\mu = \frac{s}{r_0}$ as the corresponding dimensionless parameter. We define $r_0$ and $L_0$ as the cylinder radius and length in the undeformed state, $r$ and $L_c$ as the cylinder radius and length, and $\kappa = \frac{1}{R}$ as the curvature of the center line in the deformed state, as shown in Fig. S1. Then the metric tensor $g$ equals the identity matrix,

$$g = \begin{pmatrix} 1 & 0 \\ 0 & 1 \end{pmatrix} \; .$$

Notice the constraint that the straight reinforced line on the tube has a fixed length. By using geometric relations, we have the expression for the strain tensor:

$$E^{\alpha\beta} = E_{\alpha\beta} = \frac{1}{2}(G_{\alpha\beta} - g_{\alpha\beta})$$
$$= \frac{1}{2} \begin{pmatrix} \left(\frac{r}{r_0}\right)^2 - 1 & 0 \\ 0 & \left(\frac{R+r\cos\mu}{R-r}\right)^2 - 1 \end{pmatrix} \; . \tag{14}$$

The off-diagonal elements are zero since there is no shear in this deformation. The diagonal term indicates how much the material is stretched in the circumferential and longitudinal directions. We define $\lambda_\alpha$ as the stretching rate in

corresponding directions [12]. This number $\lambda$ equals 1 when there is no stretching, bigger than 1 for actual stretching and smaller than 1 for compression. Thus we have

$$\lambda_s = \frac{|d\boldsymbol{r'}_s|}{|dr_s|} = \sqrt{1 + 2E_{ss}} = \frac{r(p)}{r_0} \ ,$$
$$\lambda_z = \frac{|d\boldsymbol{r'}_s|}{|dr_z|} = \sqrt{1 + 2E_{zz}} = \frac{R + r\cos\mu}{R - r} \ . \tag{15}$$

Notice that under the approximation of small strain $|E_{\alpha\alpha} - 1| \ll 1$, the rate of stretching becomes $\lambda_\alpha - 1 \approx E_{\alpha\alpha}$.

Consider the elastic energy of one single spring, $\Delta E = \frac{1}{2}k\Delta l^2$. Using the stretching rate $\lambda$ and assuming that the spring lies along the stretching direction, we can rewrite the energy expression: $\Delta E = \frac{1}{2}kl^2(\lambda - 1)^2$, in which $l$ in the length of the spring in the undeformed state. We compare this to the general elastic energy expression we have (Eq. 12). Thus instead of using $E_{\alpha\beta} = (\lambda^2 - 1)/2$, we define $\hat{E}_{\alpha\beta} = \lambda - 1$. This goes back to the linear elasticity model in the region of small strain. Then we have

$$\hat{E}^{\alpha\beta} = \hat{E}_{\alpha\beta} = \begin{pmatrix} \frac{r}{r_0} - 1 & 0 \\ 0 & \frac{R + r\cos\mu}{R - r} - 1 \end{pmatrix} \ . \tag{16}$$

We are now in the position to rewrite the general energy expression using $r(p)$ and $Lc(p)$ and minimize the total energy $U = \Phi - pV$ with regard to the two parameters. Defining dimensionless parameters $\bar{p} = \frac{pr_0}{Et}$, $\bar{U} = \frac{2U}{\pi r_0 L_0 Et}$, $\alpha = \frac{r}{r_0}$ and $\beta = \frac{L_c}{L_0}$ we have:

$$\bar{U} = \frac{2}{1 - \nu^2}(\alpha - 1)^2 + \frac{3}{1 - \nu^2}(\beta - 1)^2$$
$$+ \frac{4\nu}{1 - \nu^2}(\alpha - 1)(\beta - 1) - 2\bar{p}\alpha^2\beta \ . \tag{17}$$

Energy minimization gives us:

$$\frac{\partial \bar{U}}{\partial \alpha} = \frac{4(\alpha - 1)}{1 - \nu^2} + \frac{4\nu(\beta - 1)}{1 - \nu^2} - 4\bar{p}\alpha\beta = 0 \ ,$$
$$\frac{\partial \bar{U}}{\partial \beta} = \frac{6(\beta - 1)}{1 - \nu^2} + \frac{4\nu(\alpha - 1)}{1 - \nu^2} - 2\bar{p}\alpha^2 = 0 \ . \tag{18}$$

This would lead to a third order equation and can be solved analytically. Here we use commercial software (Mathematica) to plot the solutions without showing the lengthy expressions. For the values of $\nu$ and $Et$, the mapping between a continuous material and the triangular mesh in the limit of small deformation [1] predicts:

$$\nu = \frac{1}{3}$$
$$Et = \frac{2}{\sqrt{3}}k \approx 1.155k \tag{19}$$

Then we have the results shown in Fig. S16 (a). This doesn't agree well with simulation data, indicating that we are not in the small deformation region. We find the value of $\nu$ and $Et$ that describes the non-linearity of the mesh well:

$$\nu = \frac{1}{5}$$
$$Et = 1.44k_s \tag{20}$$

The result is shown in Fig. S16 (b).

### B.  Calculation of the bending and twisting magnitude of a helix deformed from a cylinder

In this section we calculate the bending and twisting magnitude of a helix deformed from a cylinder using Darboux vector. We introduce the material frame $(\vec{d_1}, \vec{d_2}, \vec{d_3})$ in which $\vec{d_3}$ is the unit vector along the longitudinal direction and $\vec{d_1}$ and $\vec{d_2}$ are two unit vectors perpendicular to $\vec{d_3}$ and to each other. We assume that in the deformed state the

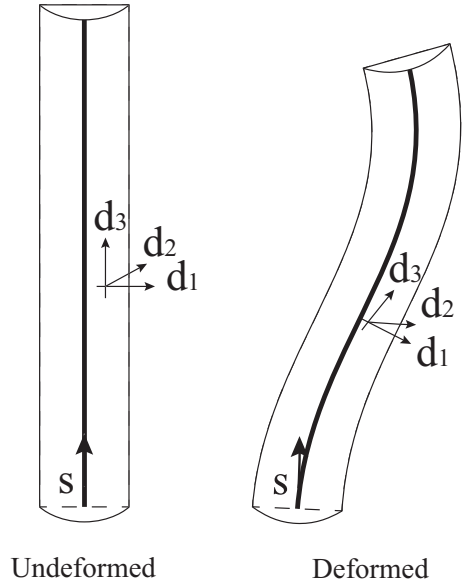

FIG. S2. The Darboux frame. Illustration of the material frame of a rod $(\vec{d_1}, \vec{d_2}, \vec{d_3})$.

three vectors are still perpendicular to each other. We use $s$ to specify the position along the central axis. Then the condition of orthonormality requires [12]:

$$\vec{d_i} \cdot \vec{d_i} = 0 \text{ and } \frac{d(\vec{d_i} \cdot \vec{d_j})}{ds} = 0 . \tag{21}$$

Therefore, there exists three scalar functions such that the derivatives of the material frame can be described as:

$$\begin{aligned}
\vec{d_1}'(s) &= \tau(s)\vec{d_2}(s) - \kappa_2(s)\vec{d_3}(s) , \\
\vec{d_2}'(s) &= -\tau(s)\vec{d_1}(s) + \kappa_1(s)\vec{d_3}(s) , \\
\vec{d_3}'(s) &= \kappa_2(s)\vec{d_1}(s) - \kappa_1(s)\vec{d_2}(s) .
\end{aligned} \tag{22}$$

Here prime ($'$) denotes the derivation with regards to $s$. We define the Darboux vector $\Omega(s)$ as:

$$\vec{\Omega}(s) = \kappa_1(s)\vec{d_1}(s) + \kappa_2(s)\vec{d_2}(s) + \tau(s)\vec{d_3}(s) , \tag{23}$$

so that Eq. 22 can be rewritten as

$$\vec{d_i}'(s) = \vec{\Omega}(s) \times \vec{d_i}(s) . \tag{24}$$

In the following we want to calculate the Darboux vector $\vec{\Omega}$ of a general case in which a tube with a reinforced area specified by $\alpha$ into a helix with known shape. We utilize the assumption that the reinforced area lies at the inner most part of the helix.

We begin with the parameterization of the surface of a helix. The expression for a helix with parameterization $(\theta, \phi)$ shown in Fig. S3 (c) is

$$\vec{r}(\theta, \phi) = ((R + r\cos\phi)\cos\theta + \frac{Lr\sin\theta\sin\phi}{\sqrt{L^2 + R^2}}, (R + r\cos\phi)\sin\theta - \frac{Lr\cos\theta\sin\phi}{\sqrt{L^2 + R^2}}, L\theta + \frac{Rr\sin\phi}{\sqrt{L^2 + R^2}}) . \tag{25}$$

We first express the material frame $(\vec{d_1}, \vec{d_2}, \vec{d_3})$. Since $\vec{d_3}(s)$ is the unit vector along the central axis at position $s$, we express the function for the central axis of the helix:

$$\vec{x}(s) = (R\cos\theta, R\sin\theta, L\theta) , \text{ with } s = \theta\sqrt{R^2 + L^2} . \tag{26}$$

Thus we have the expression for $\vec{d_3}(s)$:

$$\vec{d_3} = \frac{d\vec{x}(s)}{ds} = \frac{R}{\sqrt{R^2 + L^2}}(-\sin\theta, \cos\theta, L/R) = (-\cos\zeta\sin\theta, \cos\zeta\cos\theta, \sin\zeta) . \tag{27}$$

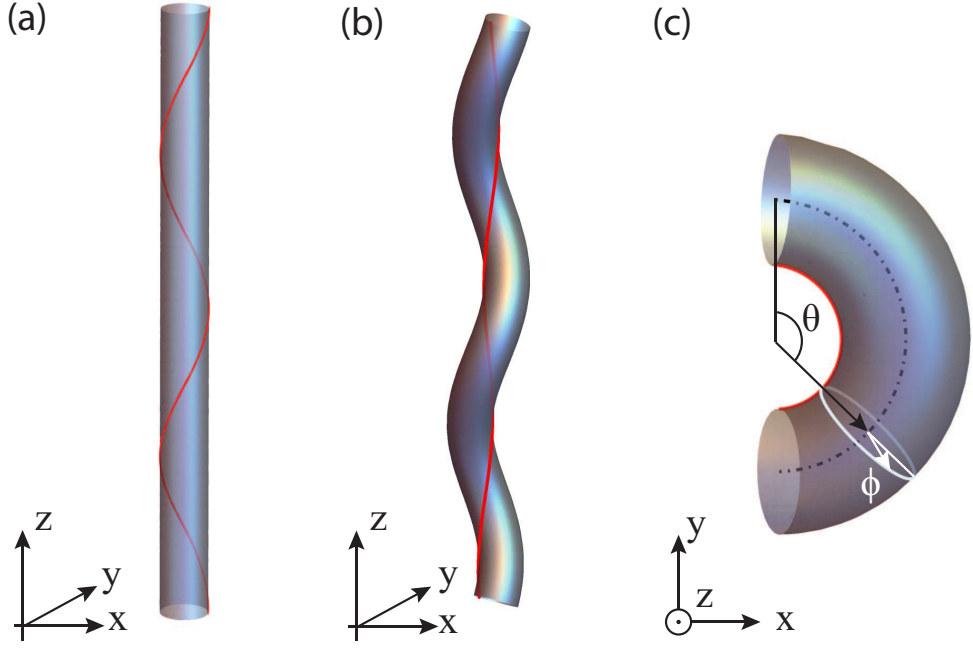

FIG. S3. Reinforced line considering different tube configurations. a) Illustration of the tube with the reinforced line in the undeformed state. b) Illustration of the tube in the shape of a helix with reinforced line (deformed state). c) Illustration showing the parameterization $(\theta, \phi)$.

For the last step, we use the definition of $\zeta$ ($\tan\zeta = L/R$).

To come up with the expression for $\vec{d}_2(s)$, we specify a straight line as a reference line on the surface of undeformed cylindrical tube and calculate the function for this line in the deformed state. The vector $\vec{d}_2(s)$ can then be expressed as the vector starting at certain postion $s$ on the central axis pointing towards the reference line on the plain locally perpendicular to the central axis. In the undeformed state, the system is a cylindrical tube. We can parameterize the undeformed state by $\vec{r}(r = r_0, R = 0, L = L_r)$, in which $\vec{r}$ is specified in Eq. 25:

$$\vec{r}_{cylinder}(\theta_1, \phi_1) = \vec{r}(r = r_0, R = 0, L = L_r) = (r_0\cos(\theta_1 - \phi_1), r_0\sin(\theta_1 - \phi_1), L_r\theta_1) \ . \tag{28}$$

Here we use the parameterization $(\theta_1, \phi_1)$ to specify the undeformed case. Thus one possible reference line expression is (the reference line is straight in the undeformed state):

$$\vec{r}_{r,1} = (r_0, 0, L_r\theta_1) = \vec{r}_{cylinder}(\theta_1, \phi_1 = \theta_1) \ . \tag{29}$$

To come up with the function of the reference line in the deformed state, we need to use the expression of the reinforced line in both the undeformed and deformed state. In the undeformed state, the expression of the reinforced line is:

$$\vec{r}_{f,1} = (r_0\cos(\theta_1), r_0\sin(\theta_1), L_r\theta_1) = \vec{r}_{cylinder}(\theta_1, \phi_1 = 0) \ . \tag{30}$$

In the deformed state, the system becomes a helix. According to the parameterization $(\theta, \phi)$ illustrated in Fig. S3 (c), $\phi = 0$ always corresponds to the outer-most part of the helix. Since we assume that the reinforced line always lies in the inner-most part of the helix, we can use $\phi_2 = \pi$ to describe the position of the reinforced line:

$$\vec{r}_{f,2} = \vec{r}(\theta_2, \phi_2 = \pi) \ . \tag{31}$$

Here we use the parameterization $(\theta_2, \phi_2)$ to specify the deformed (helix) case. In the modeling, we only consider three kinds of deformation: bending, twisting and uniform swelling. For the process of bending and twisting, given two points on the same cross-section, we know the difference in $\phi$ between them is preserved during deformation. This can be seen by noticing that any cross-section of the system remains undeformed during bending or twisting (or combined). These two kinds of deformation change the relative positioning between the cross-sections, but don't deform the cross-sections themselves. For the uniform swelling, the tube radius and contour length increase in a

uniformed way. Thus the difference in angle $\phi$ for two points on the same cross-section is still preserved. Thus we know the function of the reference line is:

$$\vec{r}_{r,2} = \vec{r}(\theta_2, \phi_2 = \pi + \phi_1) \ . \tag{32}$$

Please note $\theta_1$ and $\theta_2$ are linerly related, where $\theta_1$ ranges from 0 to $n_r$ and $\theta_2$ ranges from 0 to $n$. Here $n_r$ is the number of turns reinforced line while $n$ is the number of turns of the tube. Thus we have

$$\theta_1 = e\theta_2 \ , \text{with } e = \frac{n_r}{n} \ . \tag{33}$$

With the function of the reinforced line at hand, we can express the vector $\vec{d}_2(s)$ as the vector starting at certain position $s$ on the central axis pointing towards the reference line on the plain locally perpendicular to the central axis.

$$\vec{d}_2(s) = \frac{1}{r} \left[ \vec{r}_{r,2} - \vec{x}(s) \right] = \frac{1}{r} \left[ \vec{r}\left(\theta_2, \phi_2 = \pi + \frac{n_r}{n}\theta_2\right) - (R\cos\theta_2, R\sin\theta_2, L\theta_2) \right] \ ,$$
$$= (\cos(\pi + e\theta_2)\cos\theta_2 + \sin\theta_2\sin(\pi + e\theta_2)\sin\zeta, \cos(\pi + e\theta_2)\sin\theta_2 - \cos\theta_2\sin(\pi + e\theta_2)\sin\zeta, \sin(\pi + e\theta_2)\cos\zeta) \ . \tag{34}$$

Again for the last step, we use the definition of $\zeta$ ($\tan\zeta = L/R$). To get $\vec{d}_1$, we use $\vec{d}_1 = \vec{d}_2(s) \times \vec{d}_3(s)$. We can then calculate $\tau$, $\kappa_1$ and $\kappa_2$ by

$$\tau = \vec{d}_1'(s) \cdot \vec{d}_2(s) = -\vec{d}_2'(s) \cdot \vec{d}_1(s) \ ,$$
$$\kappa_1 = \vec{d}_2'(s) \cdot \vec{d}_3(s) = -\vec{d}_3'(s) \cdot \vec{d}_2(s) \ , \tag{35}$$
$$\kappa_2 = -\vec{d}_1'(s) \cdot \vec{d}_3(s) = \vec{d}_3'(s) \cdot \vec{d}_1(s) \ .$$

Finally we got the results:

$$\tau = (e - \sin\zeta)\frac{1}{\sqrt{L^2 + R^2}} \ ,$$
$$\kappa_1 = -\cos\zeta\cos(e\phi)\frac{1}{\sqrt{L^2 + R^2}} \ , \tag{36}$$
$$\kappa_2 = \cos\zeta\sin(e\phi)\frac{1}{\sqrt{L^2 + R^2}} \ .$$

Please note during the calculation we use the relation $\frac{d\theta}{ds} = \frac{1}{\sqrt{L^2+R^2}}$. So we have the twisting and bending magnitude:

$$\tau = \left(\frac{n_r}{n} - \sin\zeta\right)\frac{1}{\sqrt{L^2 + R^2}} \ ,$$
$$\kappa = \sqrt{\kappa_1^2 + \kappa_2^2} = \cos\zeta\frac{1}{\sqrt{L^2 + R^2}} \ . \tag{37}$$

### C. Modeling helix deformation

Here we calculate the deformation of a cylindrical tube with helical reinforced line after pressurization. A finite helix can be defined by four parameters: $(n, L, R, r)$. First we have the assumption that the length of the reinforced area is fixed: $L_f = L_f(p = 0)$. This implies:

$$2\pi n_r \sqrt{L_r^2 + r_0^2} = 2\pi n \sqrt{L^2 + (R - r)^2} \ . \tag{38}$$

For $r(p)$ and $L_c(p)$, we can use the expression derived in Supplemental Material Sec. II(A). Under these three constraints we wish to minimize the energy:

$$\Phi = \frac{EI}{2}\left(\kappa^2 + \frac{3}{4}\tau^2\right) \ , \tag{39}$$

in which (Eq. 37)

$$\tau = \left(\frac{n_r}{n} - \sin\zeta\right)\frac{1}{\sqrt{L^2 + R^2}} \; ,$$
$$\kappa = \sqrt{\kappa_1^2 + \kappa_2^2} = \cos\zeta\frac{1}{\sqrt{L^2 + R^2}} \; . \tag{40}$$

For simplification, we define the following parameters:

$$\sqrt{\eta} = \frac{L_f}{L_c(p)} \; ,$$
$$\gamma = \frac{r(p)}{(1-\eta)L} \; ,$$
$$g = \frac{(1-\eta)\gamma}{1+\beta^2} = \frac{\beta \pm \sqrt{\eta(1+\beta^2) - 1}}{1+\beta^2} \; ,$$
$$\hat{n}_0 = \frac{2\pi n_r r(p)}{L_c(p)} \; ,$$
$$\beta = \cot(\zeta) = \frac{R}{L} \; . \tag{41}$$

Using these we have the expression for $N$ and $\beta = \cot(\zeta) = \frac{R}{L}$:

$$\beta(L) = \gamma \pm \sqrt{\gamma^2\eta - 1} = \frac{1 \pm \sqrt{\eta - g^2}}{g} \; , \tag{42}$$

$$n(L) = \frac{L_c(p)}{2\pi r}g\sqrt{1+\beta^2} \; . \tag{43}$$

We rewrite the energy expression:

$$\Phi\frac{2}{EI}\frac{L_c^2}{(2\pi)^2} = n^2 - \frac{n^2}{4(1+\beta^2)} + \frac{3n_r^2}{4} - \frac{3}{2}\frac{n}{\sqrt{1+\beta^2}}n_r \; . \tag{44}$$

Since we can express both $\beta$ and $n$ as a function of $L$, we require $\frac{d\Phi}{dL} = 0$ and get the following:

$$4g\sqrt{1+\beta^2}\left(\frac{dg}{dL}\sqrt{1+\beta^2} + g\frac{d\sqrt{1+\beta^2}}{dL}\right) - g\frac{dg}{dL} - 3\hat{n}_0\frac{dg}{dL} = 0$$

$$\Leftrightarrow 4\frac{dg}{dL}\frac{\eta + 1 \pm 2\sqrt{\eta - g^2}}{g} - 4\frac{dg}{dL}\frac{1 + \eta \pm 2\sqrt{\eta - g^2} \pm \frac{g}{\sqrt{\eta - g^2}}}{g} - g\frac{dg}{dL} - 3\hat{n}_0\frac{dg}{dL} = 0$$

$$\Leftrightarrow \pm\frac{4g}{\sqrt{\eta - g^2}} + g + 3\hat{n}_0 = 0 \tag{45}$$

$$\Leftrightarrow 16g^2 = (g + 3\hat{n}_0)^2(\eta - g^2)$$

$$\Leftrightarrow g^4 + 6\hat{n}_0 g^3 + (16 - \eta + 9\hat{n}_0^2)g^2 - 6\hat{n}_0\eta g - 9\hat{n}_0^2\eta = 0 \; .$$

Thus we can solve for $g$ at different pressures and number of turns $n_0$. This would also give us the values of $\beta$ and $n$. We can derive the expression $L = \frac{L_c(p)}{2\pi n\sqrt{1+\beta^2}}$ to calculate the values of $L$. Recall $R = L\beta$. Thus we solve the system of $(n, L, R, r)$. Throughout the parameter region considered, the equation for $g$ only has one positive real root so there is no ambiguity. We select the minus sign in the $\beta(g)$ expression (Eq. 42). A comparison between the results of simulation and analytical formulation can be found in Fig. S9 and Fig. 2 in the main text.

### D. Critical pressure for the maximum helix height

Due to the constraint that the reinforced path is not stretchable, we have:

$$L_f = \sqrt{z^2 + (2\pi n)^2 \cdot (R - r)^2} \; . \tag{46}$$

Thus the maximum of $z$ is achieved when $R = r$. On the other hand, we have the expression for $R$:

$$R = \beta L = \beta \cdot \frac{L_c}{2\pi n \sqrt{1 + \beta^2}} = \frac{r}{g(\beta + 1/\beta)} \ , \tag{47}$$

in which

$$\beta = \frac{1 - \sqrt{\eta - g^2}}{g} \ . \tag{48}$$

Thus $R = r$ gives us

$$\eta^2 = \eta - g \ . \tag{49}$$

On the other hand, the equation for $g$ (Eq. 45) gives us

$$16g^2 = (g + 3\hat{n}_0^2)^2(\eta - g) \ . \tag{50}$$

Thus we can solve the $g$ and $\eta$ for the critical pressure of maximum height. Recall the definition $\eta = \frac{L_f^2}{L_c^2}$, which implies $\eta$ is close to 1. We can assume $\eta = 1 - \varepsilon$ with $|\varepsilon|$ a small parameter, and solve for $\hat{n}_0$ up to the first order of $\varepsilon$. Thus Eq. 49 becomes $g = \varepsilon$. Bring this relation to Eq. 50 and neglect the higher order terms of $\varepsilon$, we have:

$$\hat{n}_0^2 = \varepsilon = 1 - \eta = 1 - \frac{L_f^2}{L_c^2} \ . \tag{51}$$

Recall the definition $\hat{n}_0 = \frac{2\pi n_r r(p)}{L_c(p)}$ (Eq. 41), we have:

$$(2\pi n_r r)^2 = L_c^2 - L_f^2$$
$$\Leftrightarrow L_{c0}^2 \cot^2(\alpha) \frac{r^2}{r_0^2} = L_c^2 - L_{c0}^2(1 + \cot^2(\alpha)) \ . \tag{52}$$

So we have the final result as shown in the main text:

$$\cot^2(\alpha) = \frac{(L_c/L_{c0})^2 - 1}{(r/r_0)^2 + 1} \ . \tag{53}$$

Here $L_{c0}$ is the contour length in the undeformed state, which equals to the helix height in the undeformed state. The right hand side depends on the material properties, which turns out to be nonlinear in our case of triangular mesh. Eq. 53 also reveals that when $\alpha$ decreases, the increase of the left hand side ($\cot^2(\alpha)$) would in general make it harder for the right hand side (which equals zero in the zero pressure case) to match the left hand side. Thus the state of maximum height may not be achievable when $\alpha$ is too small for a limited range of pressure, as observed in the simulation. With the empirical $r(p)$ and $L_c(p)$ relation provided in Supplemental Material Sec. II(A), we can calculate the range of $\alpha$ that the maximum height would be achieved during a limited range of pressure. For the current pressure range we consider ($p \leq 1.5$ atm), the maximum height would be achieved only if $\alpha > 61°$. We compare the critical pressure for the maximum height we got from Eq. 53 and from numerically calculating the critical pressure for the height predicted in this analytical framework (Fig. S4). For the large $\alpha$ region with $\alpha > 70°$, the two agrees well with each other, thus validating the assumptions we used. As the pressure increases, the contour length $L_c$ increases while the length of the reinforced line $L_f$ is fixed. Thus $|\varepsilon|$ increases as the pressure increases, making our assumption less accurate. This leads to the divergence of the two curves as pressure increases. The comparison between the theoretical prediction (Fig. S4) and the simulation result can be found in Fig. 4 in the main text.

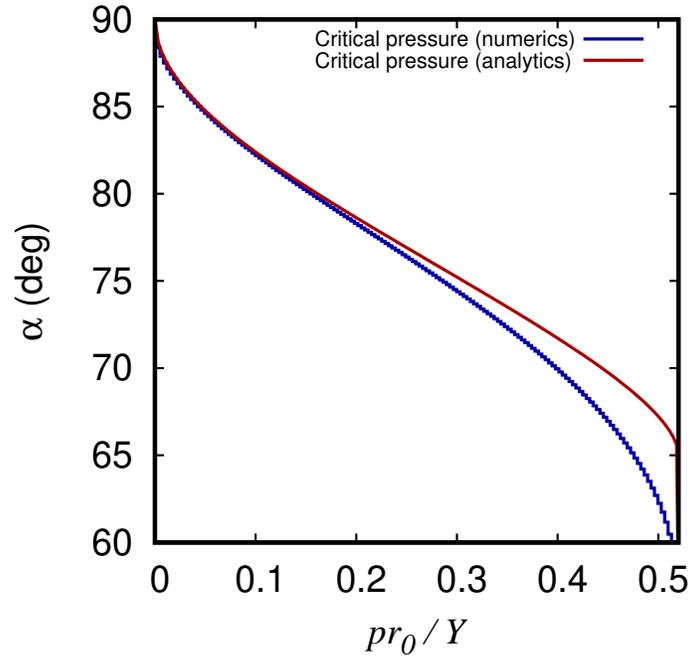

FIG. S4. Critical pressure as a function of the helical angle. Comparison between analytical results (Eq. 53) and numerical results (numerical calculation under the framework described in Supplemental Material Sec. II(C)) of the critical pressure for different values of $\alpha$.

### III. EXPERIMENTAL PROCEDURES

To create the rubber tubes, we print two mold parts using a 3D printer (Ultimmaker 3 extended) and a PLA substance. The mold is a cylindrical cavity (30 cm length, 2 cm diameter) to which a solid tube with smaller diameter is inserted. The solid tube is centered by the cap and the pin at the tip (Fig. S5), in order to keep a fixed distance (1 mm) between the mold and the tube across the structure. The solid tube has a helical engraving with a specific pitch angle (ranging 25-85°), that supports the coiling of a red embroidery thread (1 mm diameter). The thread is attached at its ends to the solid tube using scotch tape.

A liquid composition of tin-cured silicone rubber (Mold-Max NV-14, Ceramet, prepared according to instructions) is poured onto the two mold halves. The solid tube is placed on top of one of them, aligning pins and pressure clamps are used to seal the other half on top of it. The composition is left to cure overnight in a vertical position. After releasing the cured rubber tube from the mold, further sealing of the hole at the tip is performed with the same silicone rubber. Marks on the rest configuration of the rubber tube are made using a Sharpie pen.

The tube is then connected to a pressure source with a fine pressure regulator (IR2010 0.1-4bar, Baccara.co.il). Videos and images of the different configurations are taken.

Helical properties were obtained from snapshots at different pressures using the ImageJ software.

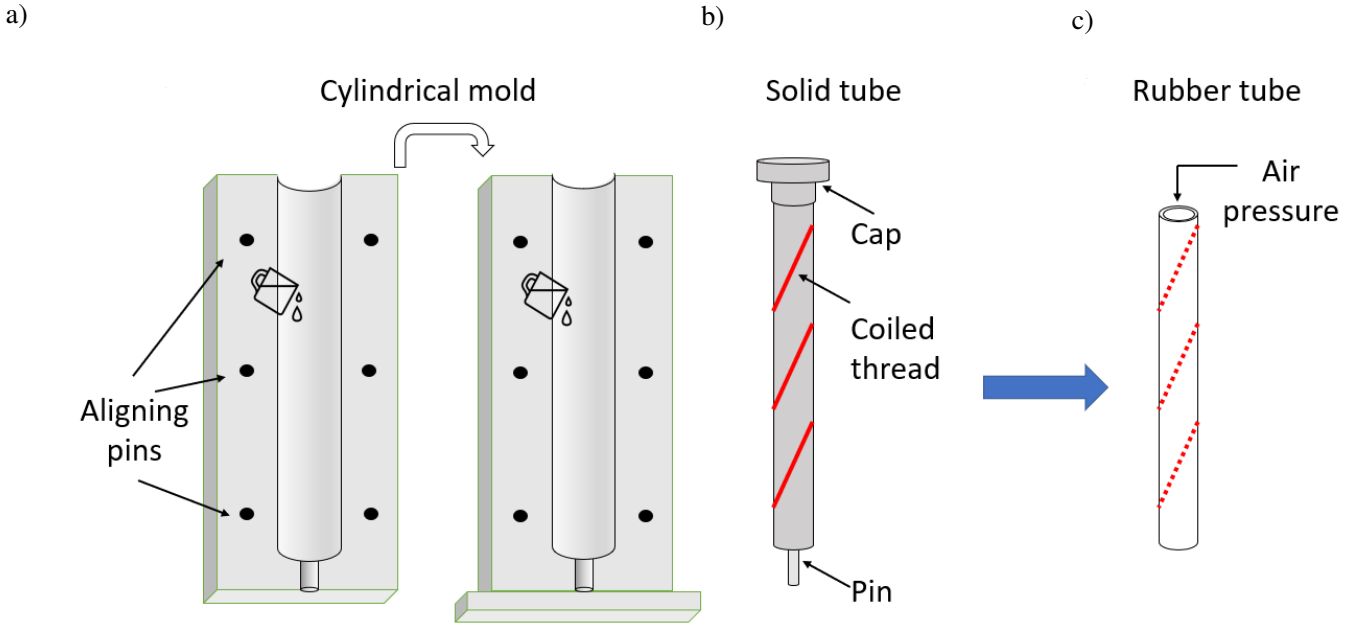

FIG. S5. Making of inflatable rubber tubes. a) The two halves of the 3D printed mold create a cylindrical cavity, to which the liquid rubber is poured. b) a 3D printed solid tube, with an unextensible thread coiled around it, is placed between the two mold halves. c) After over-night curing, the rubber tube with an embedded helical thread is released from the mold. Connected to a pressure source, it can now be inflated in a controlled manner.

## IV.   VARIATION OF OSMOLARITY IN THE STOMACH

It has been shown the osmolarity in the stomach of dogs increases by one order of magnitude after chemical-induction of gastric juice secretion, reaching the steady state ≈20 minutes post stimulation [15]. Later, [16] showed that the osmolarity 30 and 90 minutes post injections is similar, suggesting a rapid change after injestion.

In addition, [16] showed that the osmolarity is highly-dependent on the ingested food. Most of the results shown by the authors concern the analysis of gastric juices after a steak meal, finding an osmolality on the stomach content of 232 mosm./kg. The authors, with the objective of testing the role of an hypertonic meal, conducted the same measurements after a milk and doughnut meal, finding an osmolarity almost two times larger (440 mosm./kg). A similar result was reported later by [17], where a wide range of osmolalities is observed in the stomach depending on the ionic strength of a drank solution, from 20 mosm./kg for water to 360 mosm./kg for an hypertonic beverage.

# V. SUPPLEMENTARY FIGURES

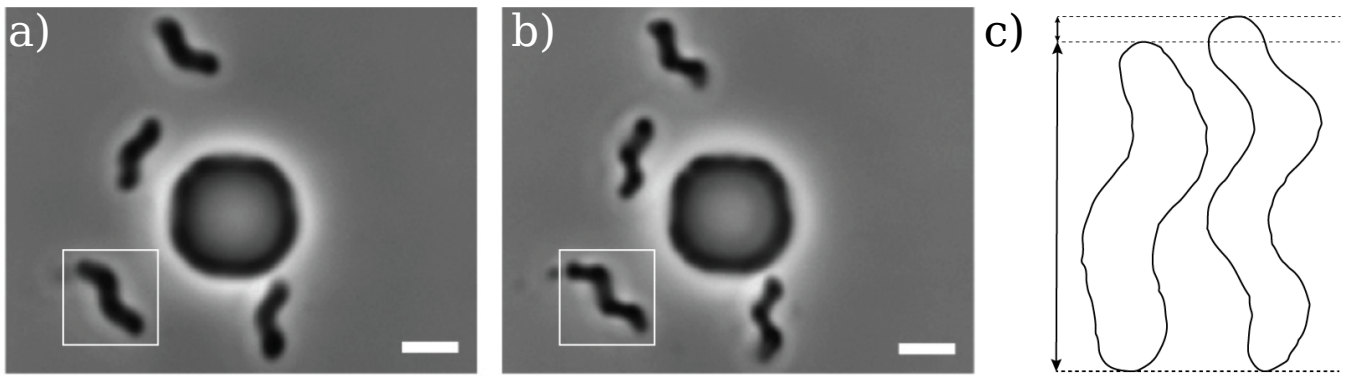

FIG. S6. Wild-type *H. pylori* cells a) before and b) after hyper-osmotic shocks with 200 mM NaCl. Figure adapted from [18]. c) The contours of one cell before and after hyper-osmotic shocks. The selected cell is labelled by the white square in panel a) and b). Note the decrease in radius and the lengthening (increase in height) after the osmotic shock.

a) $\alpha = 87.1$ deg

b) $\alpha = 85.7$ deg

c) $\alpha = 84.3$ deg

d) $\alpha = 78.6$ deg

e) $\alpha = 73.2$ deg

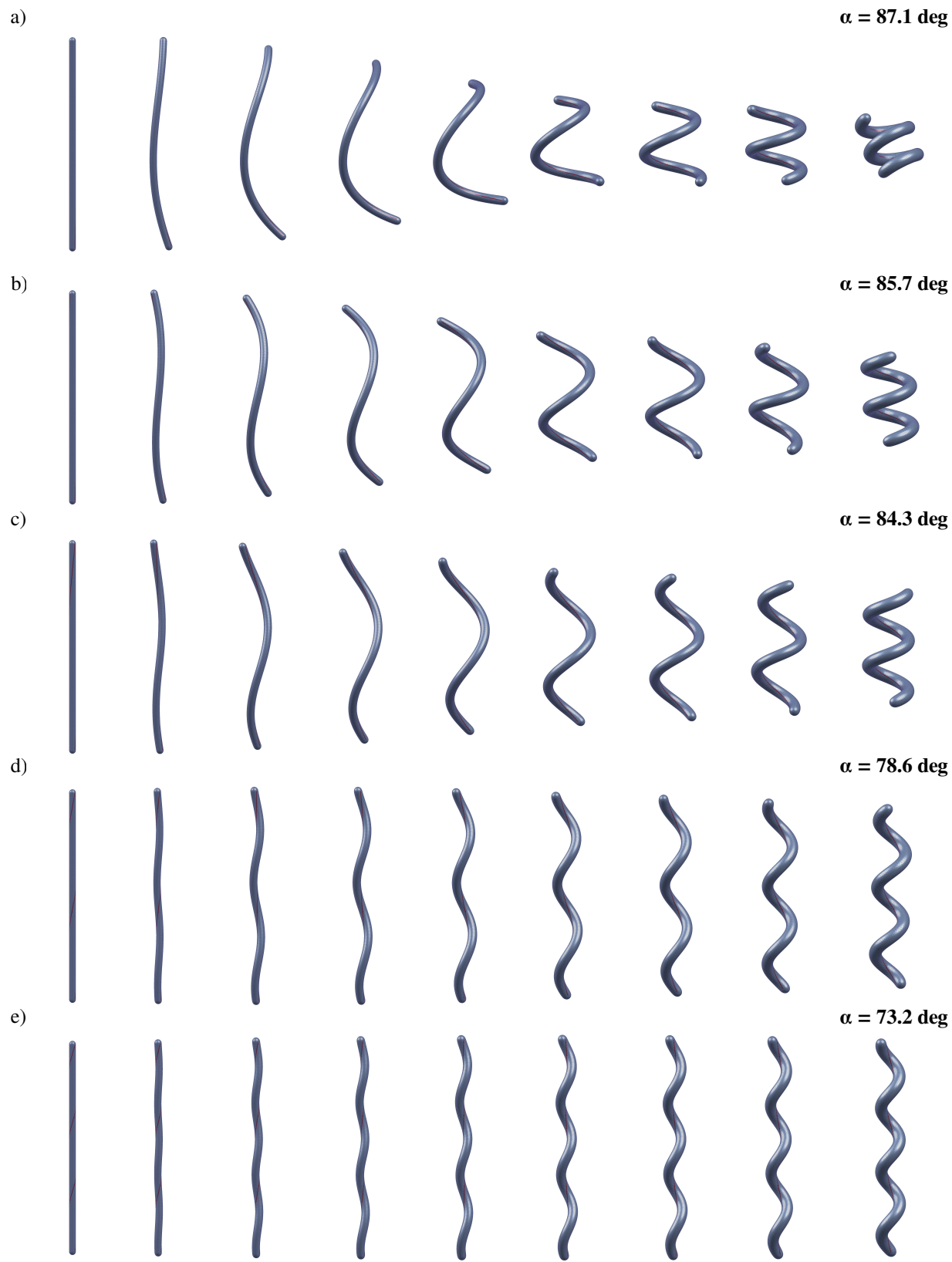

f) **α = 68.1 deg**

g) **α = 60.0 deg**

h) **α = 44.9 deg**

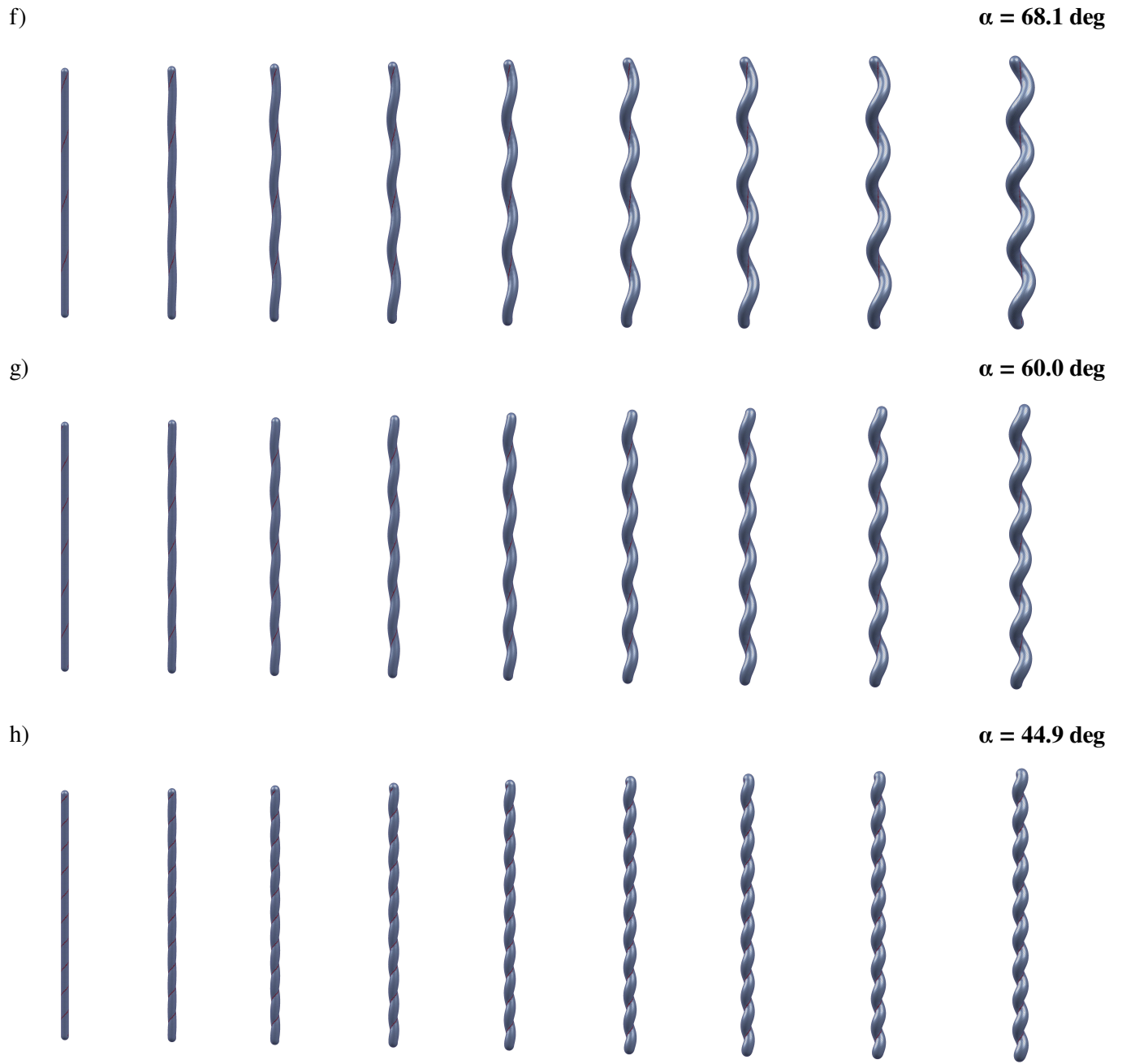

FIG. S7. Example minimal energy configurations for pressurized spherocylinders. Data corresponding to different reinforcing angles $\alpha$ for increasing rescaled pressures $\bar{p} = pr_0/Y$. The initial configuration as well as the mechanical properties of the cell are the same for every simulation. The reinforced regions is shown in red. Non-dimensional pressures, from left to right, $\bar{p} = 0$, 0.10, 0.20, 0.27, 0.34, 0.41, 0.44, 0.47, 0.51.

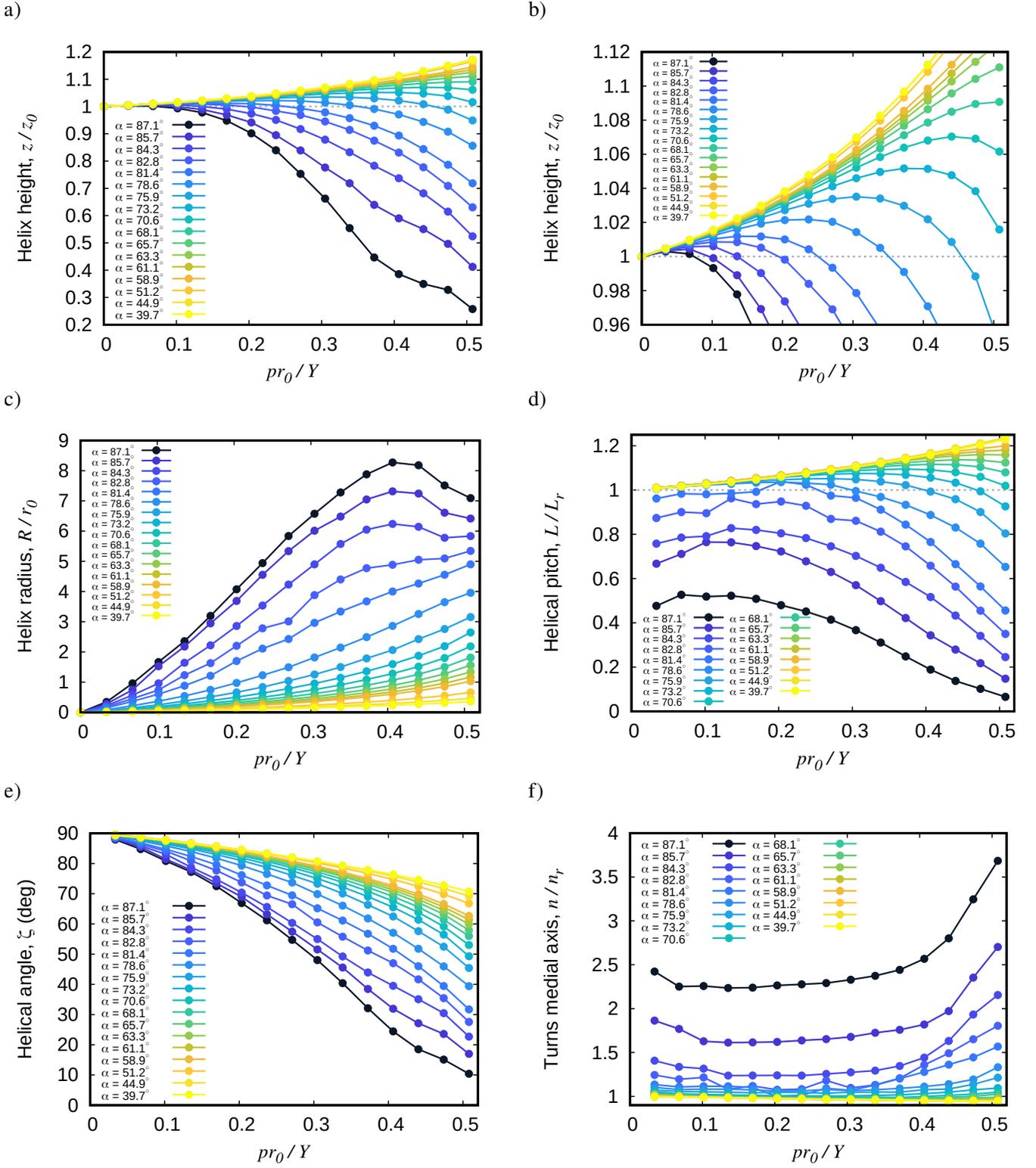

FIG. S8. Helical properties as a function of pressure and reinforcement angle. a) Helix height and b) zoomed version of showing the initial increase of height. c) Helical radius $R$. d) Helical pitch $L$ normalized by the undeformed configuration ($L_r$). e) Helical angle, $\zeta = \mathrm{atan}(L/R)$. f) Turns of the medial axis $n$ normalized by the initial number of turns of the reinforced string in the undeformed configuration ($n_r$).

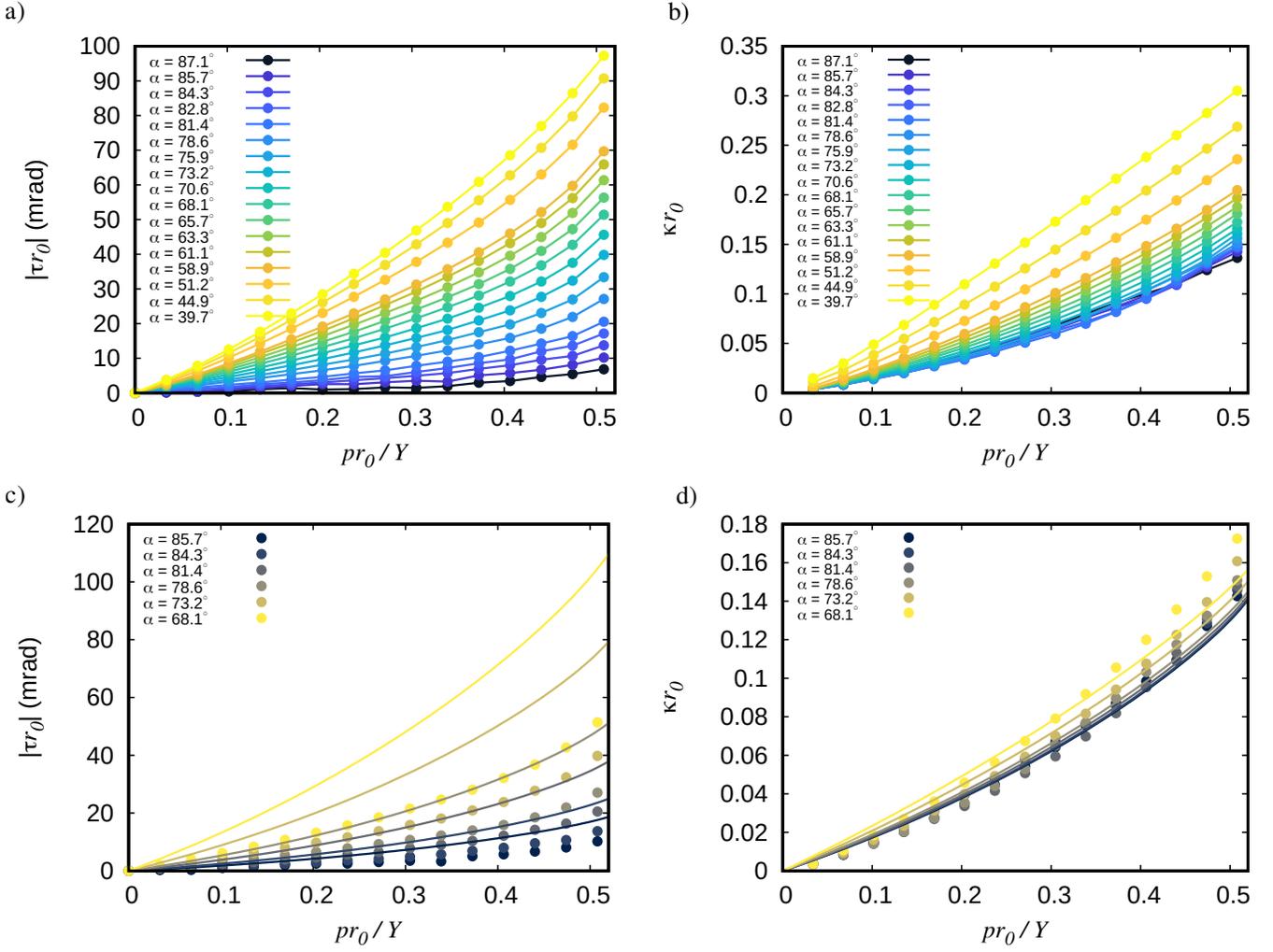

FIG. S9. Torsion and curvature dependency on pressure and reinforced angle. a) Absolute value of the torsion $\tau$. The sign of the torsion is negative (left-handed, clockwise) and its opposite to that of the reinforced helix (right-handed, counter-clockwise). b) Curvature $\kappa$ as a function of the rescaled pressure. c) Detail of the torsion $\tau$ for the steepest helical angles $\alpha$. d) Detail of the curvature $\kappa$ for the steepest angles $\alpha$. Solid lines in a) and b) are a guide to the eye. Solid lines in c) and d) are theoretical predictions following Eq. 37.

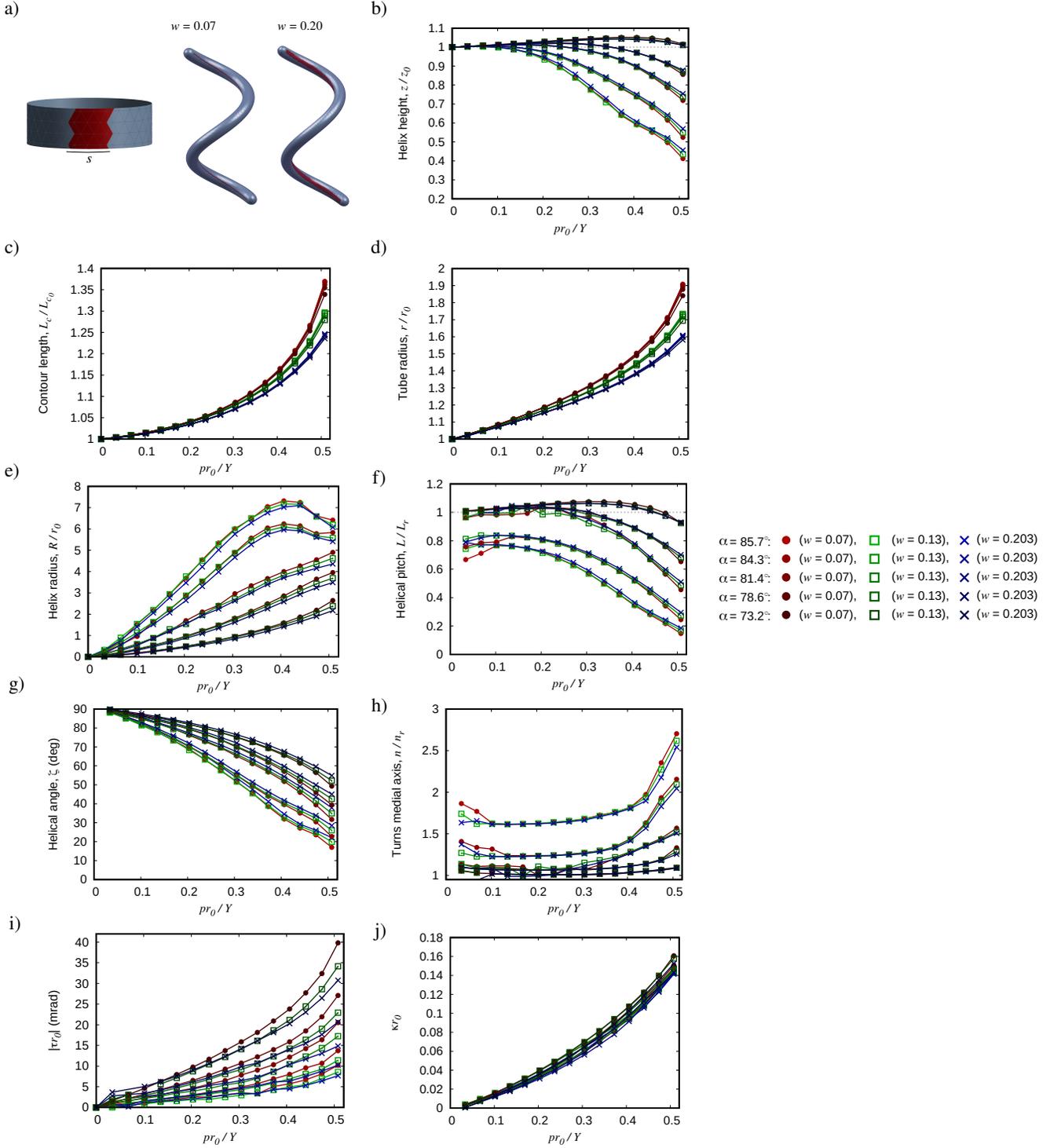

FIG. S10. Dependency of the width of reinforced string on helical parameters. a) *Left.* We define the reinforcement width $w := s/(2\pi r_0)$ as the fraction of material reinforced along the circumferential direction with respect to the perimeter of the cross-section. *Right.* Example of cells with reinforced angle $\alpha = 85.2°$ at $\bar{p} = 0.41$ for $w = 0.07$ and $w = 0.20$. b) Helix height, $z$. c) Contour length $L_c(p)$. d) Tube radius, $r(p)$. e) Helix radius, $R$ f) Helical pitch, $L$. g) Helical angle, $\zeta$. h) Number of turns of the medial axis, $n$. i) Surface torsion, $\tau$. j) Average curvature of the medial axis, $\kappa$. Dimensions and mechanical parameters are shown in Table S1.

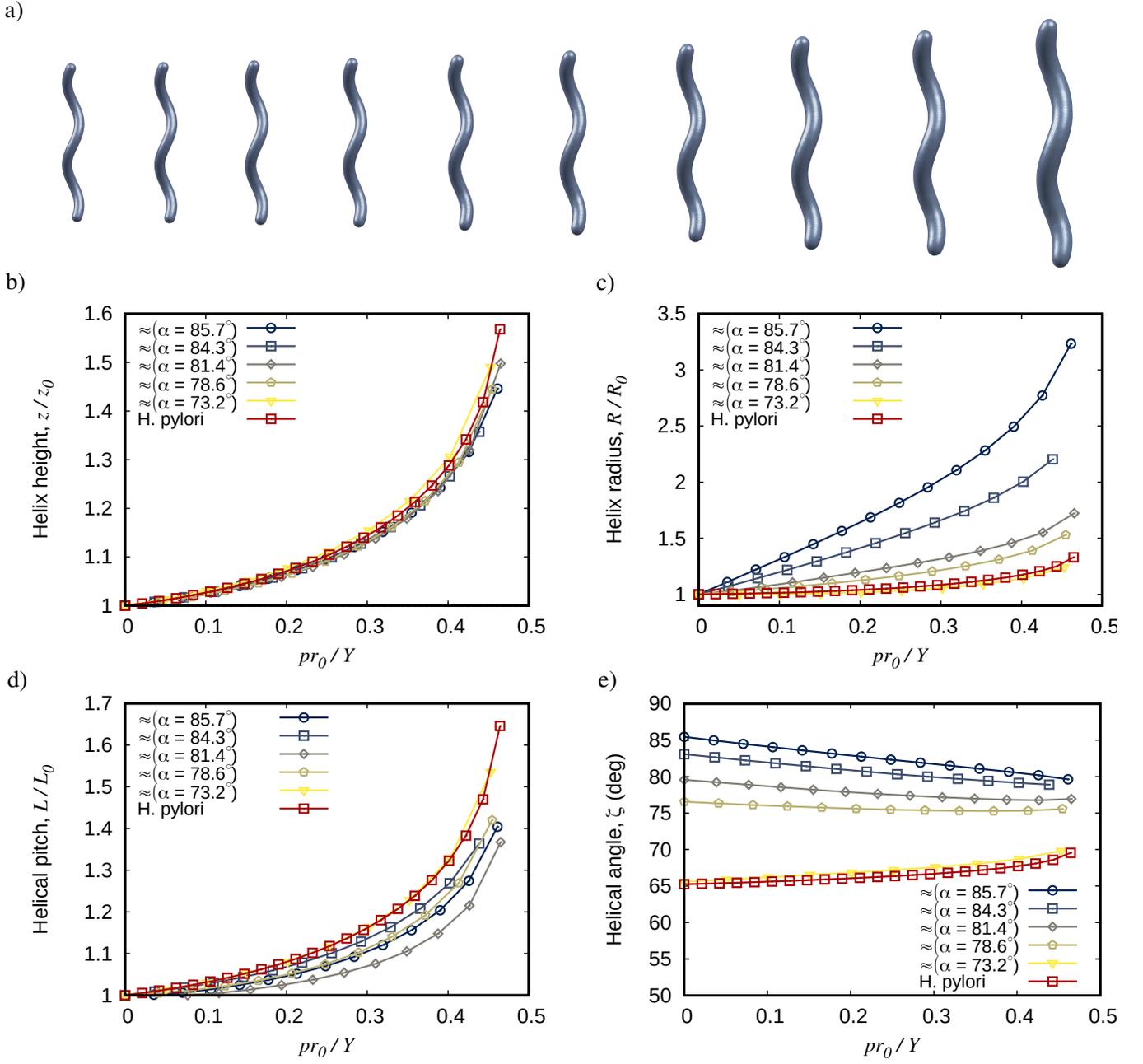

FIG. S11. a) Example minimal energy configurations for a pressurized helix. In the examples shown, the initial configuration was generated considering the mean helical parameter observed in turgid living *H. pylori* [7]. Non-dimensional pressures, from left to right, $\bar{p} = 0, 0.08, 0.15, 0.21, 0.27, 0.34, 0.40, 0.44, 0.46, 0.49$. b) Helix height, $z$. c) Helical radius, $R$. d) Helical pitch, $L$. e) Helical angle, $\zeta$. The labels $\approx (\alpha x)$ indicate an initial helical configuration with the parameters obtained at the maximum extension from the simulations of the reinforced tube with reinforcement angle $x$.

a)

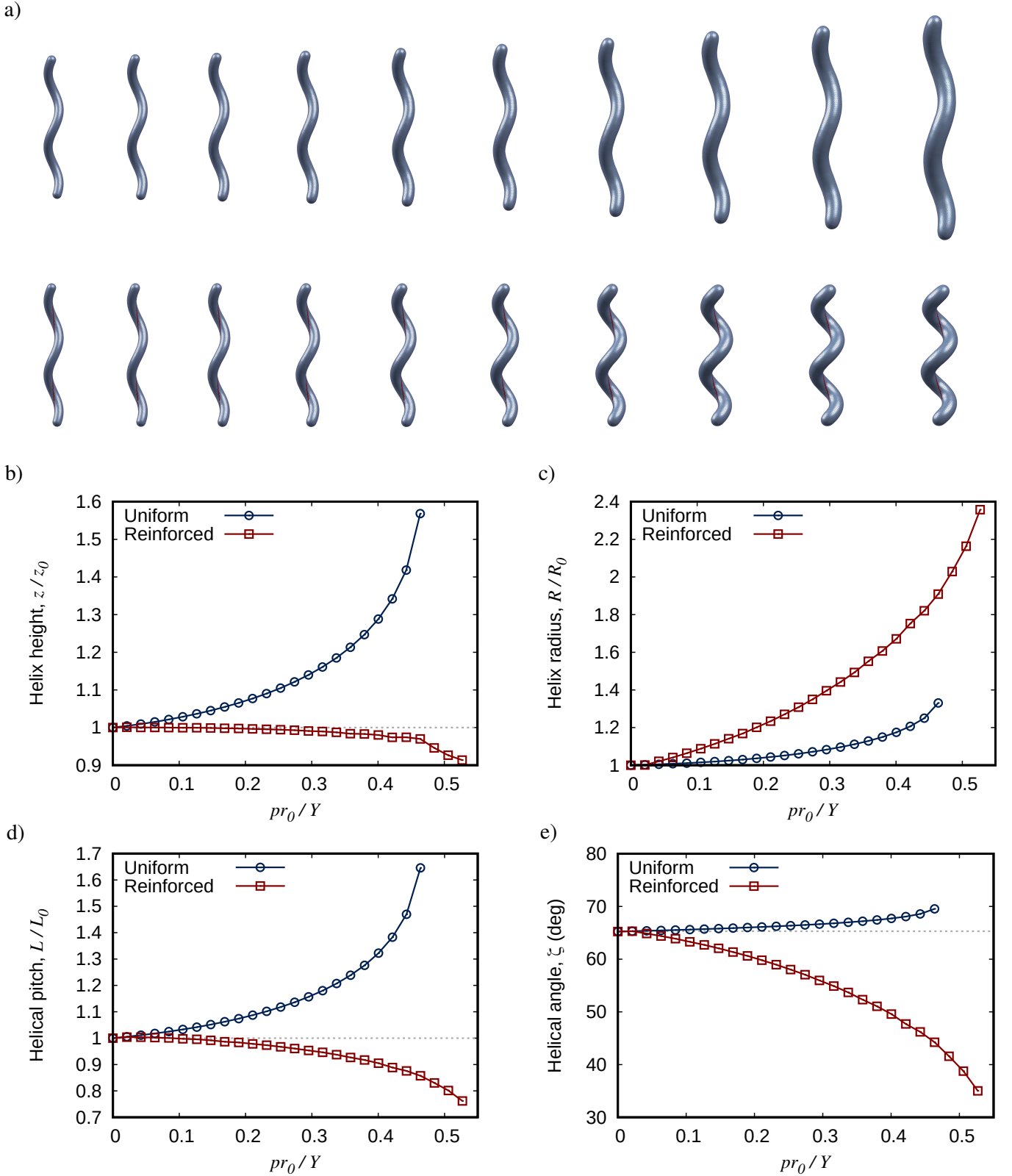

FIG. S12. a) Example minimal energy configurations in pressurized helices. *Top*: Uniform mechanics. *Bottom*: Reinforced on the concave region (red line). Non-dimensional pressures, from left to right, $\bar{p} = 0, 0.08, 0.15, 0.21, 0.27, 0.34, 0.40, 0.44, 0.46,$ 0.49. b) Helix height, $z$. c) Helical radius, $R$. d) Helical pitch, $L$. e) Helical angle, $\zeta$. For the reinforced case we obtain, at $p = 2.0$ atm ($\bar{p} = 0.42$), the following helical parameters $r = 187$ nm, $R = 263$ nm, $L = 1.8$ $\mu$m/turn and 2 helical turns. The initial configuration was generated considering the parameter observed in turgid living *H. pylori* [7].

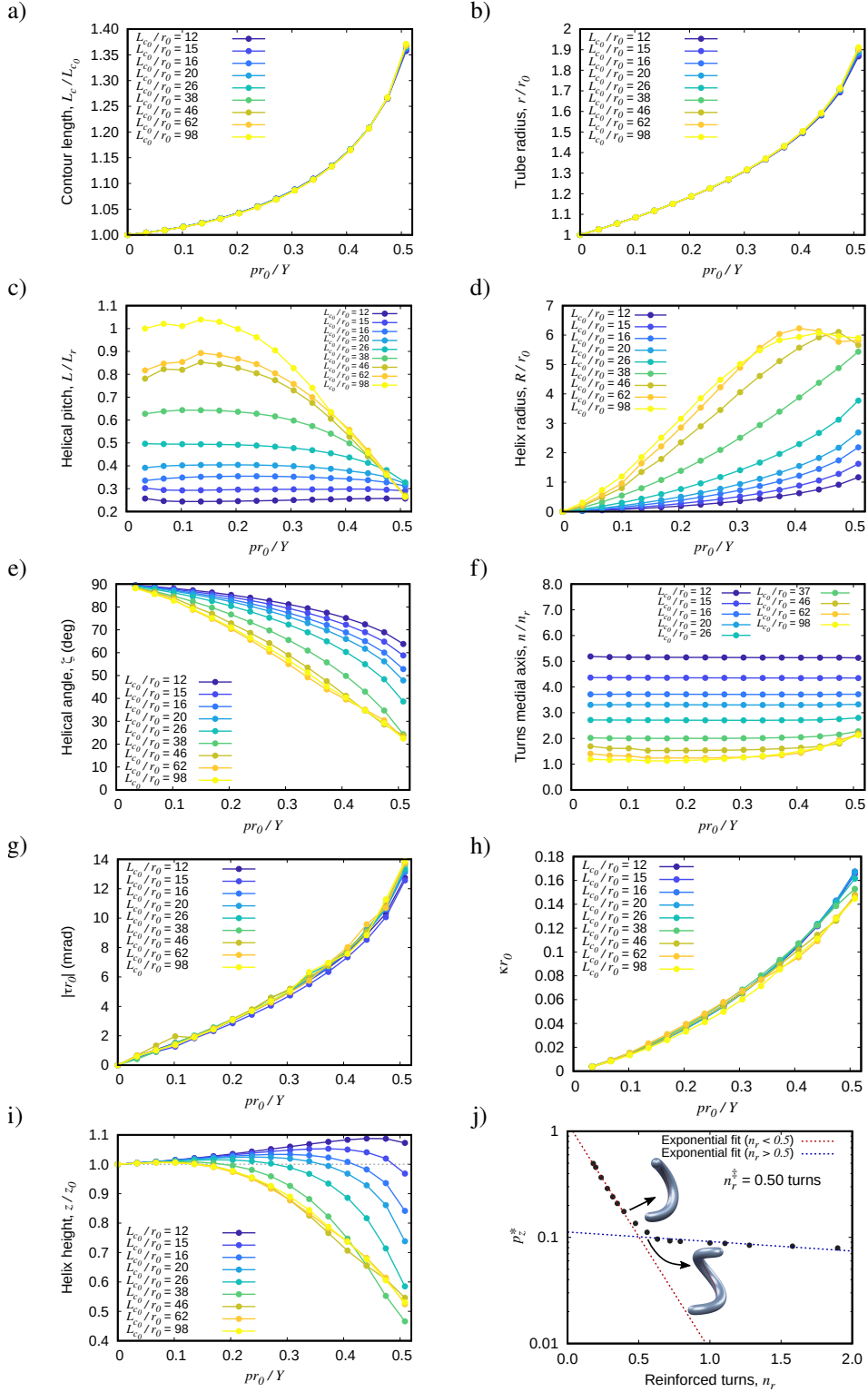

FIG. S13. Initial length and reinforced number of turns during pressure-induced formation of helices. a) Contour length $L_c(p)$. b) Tube radius, $r(p)$. c) Helical pitch, $L$. d) Helix radius, $R$. e) Helical angle, $\zeta$. f) Number of turns of the medial axis, $n$. g) Surface torsion, $\tau$. h) Average curvature of the medial axis, $\kappa$. i) Height of the main body, $z$. j) Critical pressure resulting in maximum height as a function of the initial number of reinforced turns $n_r$ (since $n_r = L_{c_0}/(2\pi r_0 \tan \alpha)$). Dashed lines are exponential fits for $n_r < 0.5$ (red) or $n_r > 0.5$. The intersection between both fits occurs at $n_r^\dagger \simeq 0.5$. *Insets*. Minimal energy configurations for $n_r = 0.40$ and $n_r = 0.56$ showing the soft transition between C-shape to helix occurring at $n_r \approx 0.5$ (($\bar{p} = 0.47$)). Data for different aspect ratios $L_{c_0}/r_0$, with constant $r_0$, for a reinforcement helical angle $\alpha = 84.3°$. The threshold $n_r = 0.5$ occurs at $L_{c_0}/r_0 = 32$. Simulation parameters are shown in Table S1.

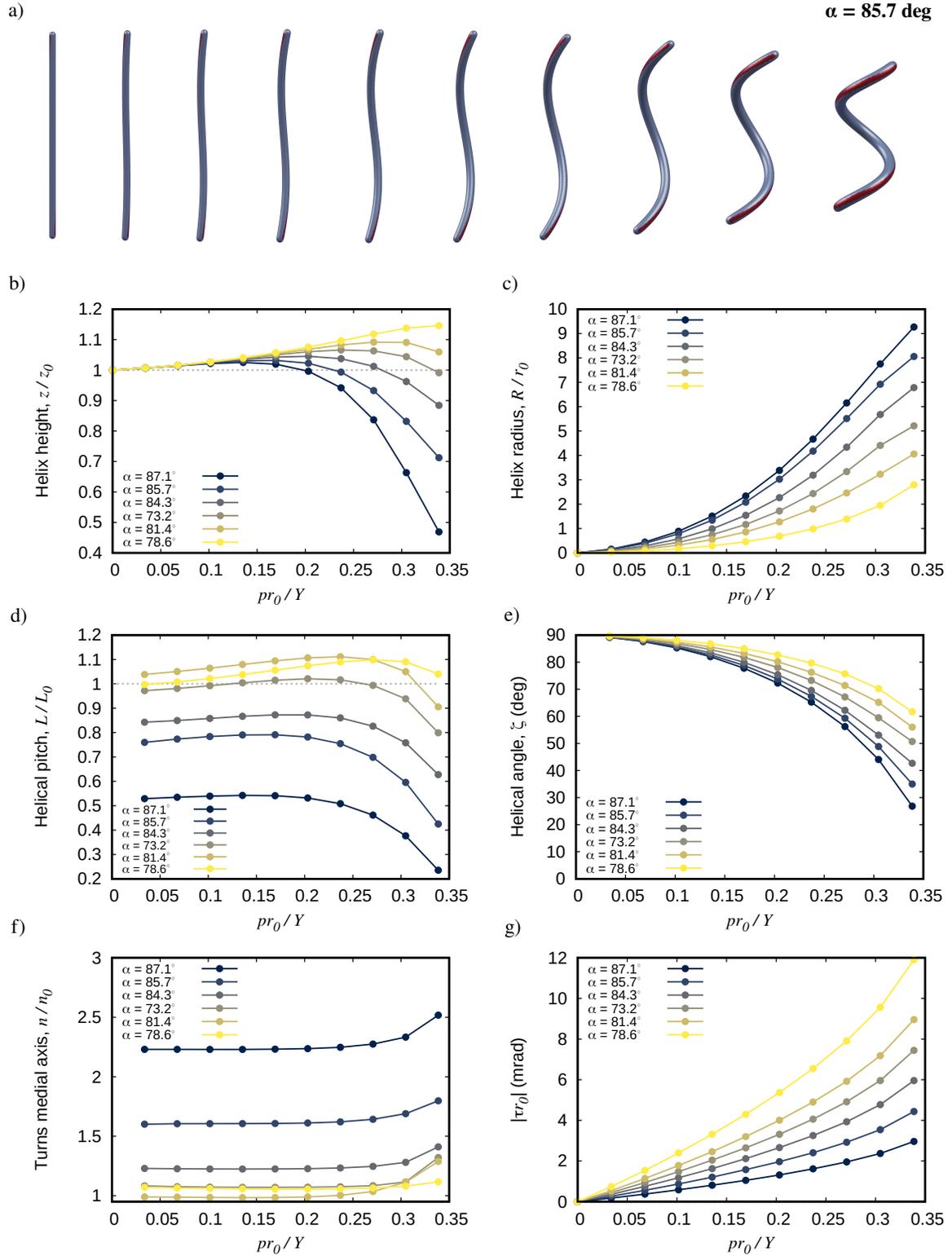

FIG. S14. Pressurization of tubes with helical weakening. a) Snapshot of minimal energy configurations for weakening angle $\alpha$ = 85.7°. After pressurization, the weakened region (red) is located on the outside (convex) area. Non-dimensional pressures, from left to right, $\bar{p}$ = 0, 0.07, 0.10, 0.14, 0.17, 0.20, 0.24, 0.27, 0.30, 0.34. b) Height of the helix $z$, c) Helical radius, $R$, d) Helical pitch, $L$, e) Helical angle, $\zeta$, f) Number of turns of the medial axis, $n$, and g) Surface torsion, $\tau$, as a function of pressure and weakening angle. The weakening is adjusted by reducing the 2D Young modulus $Y$ of the helical domain ($K$ = 0.33). Lines connecting dots are a guide to the eye.

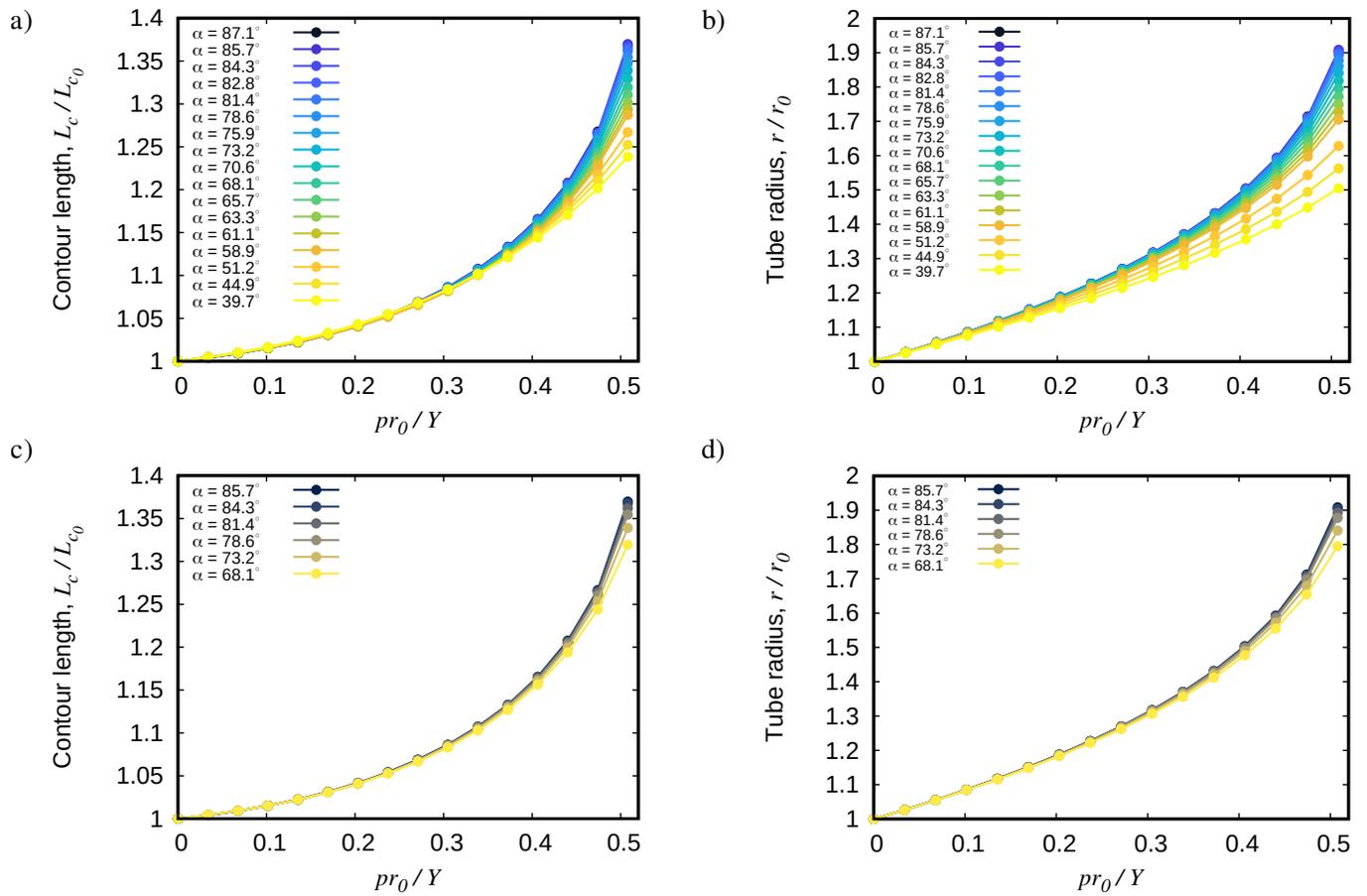

FIG. S15. Contour length and radius under pressure. Response for different reinforcement angles. Little differences are observed on $L_c(p)$ and $r(p)$ for $\alpha > 60°$.

a)

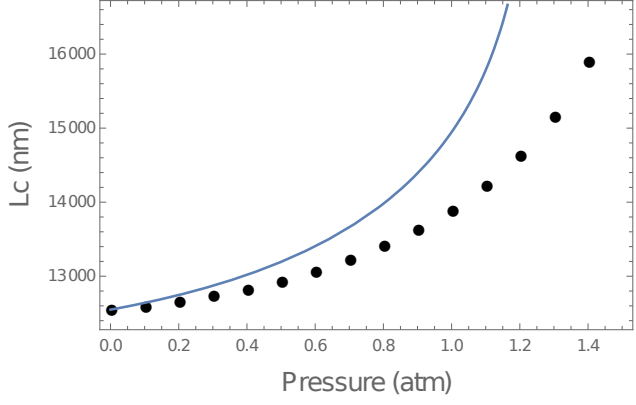
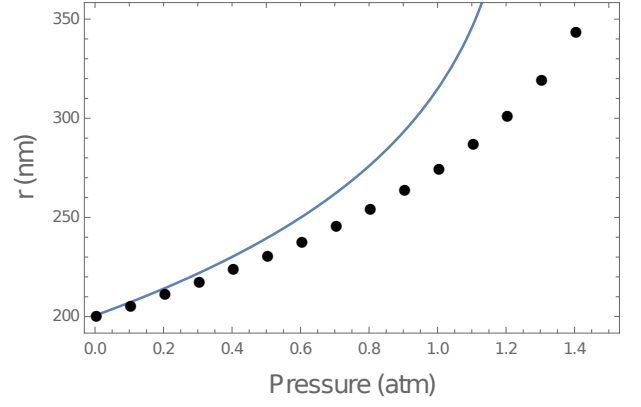

b)

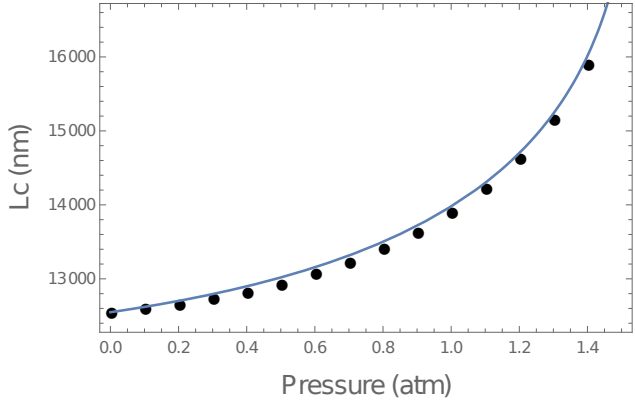
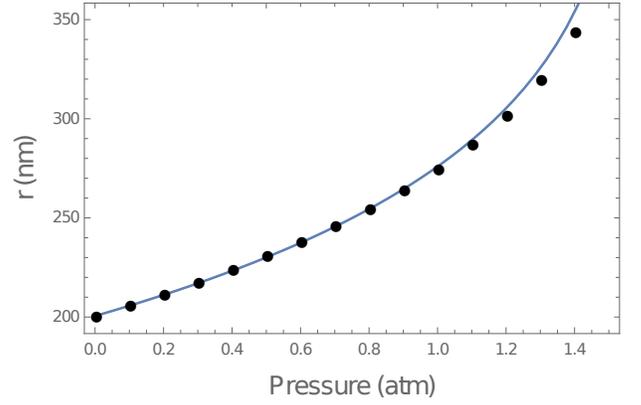

FIG. S16. Modeling of $r(p)$ and $L_c(p)$. Comparison between simulation data (dots) and modeling results (lines) of the overall swelling of the system (contour length $L_c$ and radius $r$). (a) Modeling with 2D Young modulus $Y = Et = 1.16k_s$ and Poisson's ratio $\nu = 1/3$. A mapping between a continuous material and the triangular mesh predict this relationship to hold in the limit of small deformation. (b) Modeling considering a 2D Young modulus $Y = 1.44k_s$ and Poisson's ratio $\nu = 1/5$. Notice that we are using $p$ instead of $pr_0/Y$, $L_c$ instead of $L_c/L_{c_0}$ and $r$ instead of $r/r_0$..

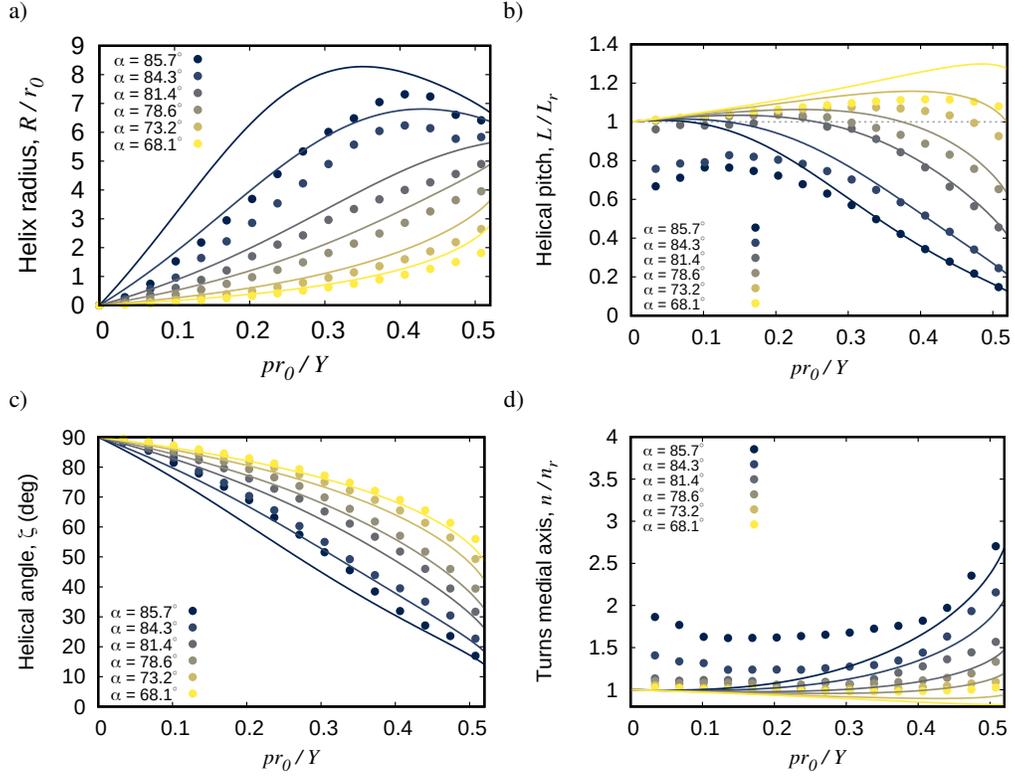

FIG. S17. Helical parameters after pressurization. a) Helical radius ($R$), b) Helical pitch ($L$) normalized by the helical pitch of the pitch of the reinforced string in the undeformed configuration $L_r$, c) Helical angle $\zeta := \mathrm{atan}(L/R)$ and d) number of turns of the medial axis $n$ normalized by the number of turns of the reinforced string in the undeformed configuration $n_r$ as a function of the rescaled pressure $\bar{p} = pr_0/Y$. Symbols are the results from numerical simulation; solid lines are model predictions.

a)

b)

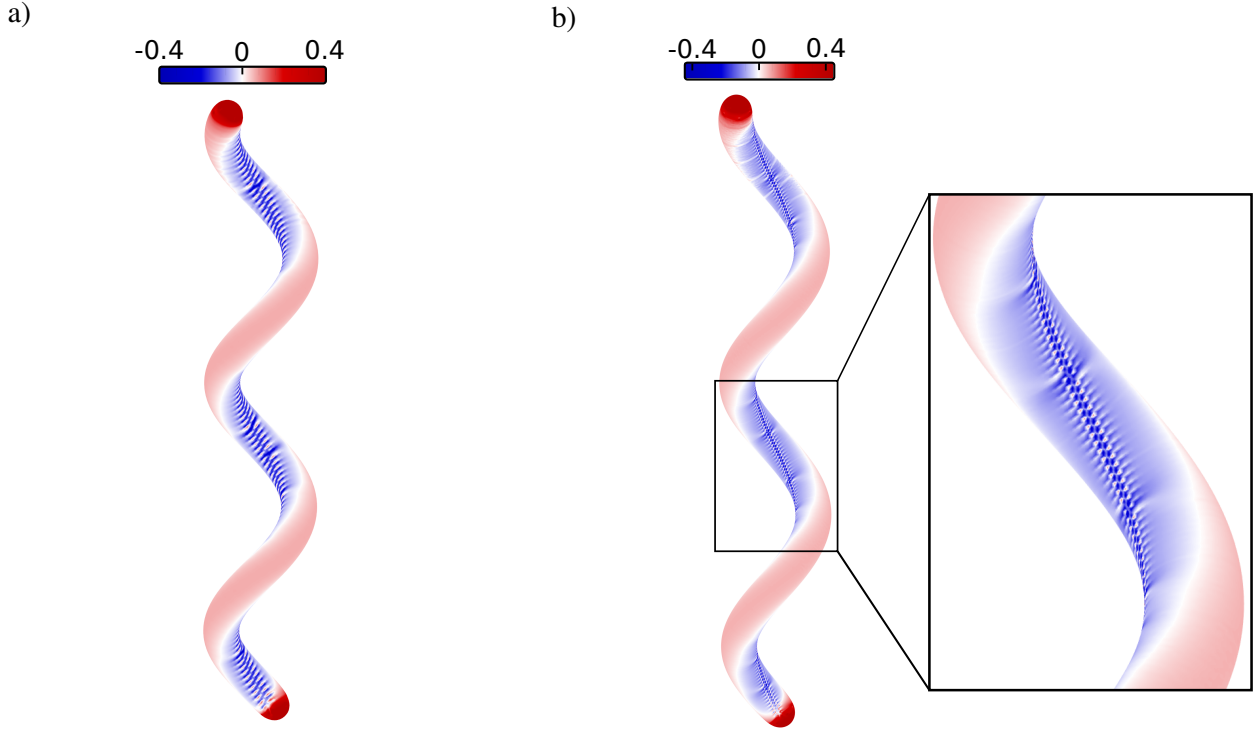

FIG. S18. Wrinkling during pressurization. a) Gaussian curvature in a pressurized helix to $\bar{p} = 0.47$ ($\alpha = 78.6°$). b) Refined mesh with $N = 35652$ vertices pressurized to $\bar{p} = 0.47$ ($\alpha = 78.6°$). Detail of the concave region highlighting the periodic wrinkling pattern. Gaussian curvature $\kappa_G$ was determined as described in [19]. The indicated values correspond to rescaled curvatures given by $\bar{\kappa_G} := \kappa_G r_0^2$. The mechanical properties and the dimensions of the helix are indicated in Table S1.

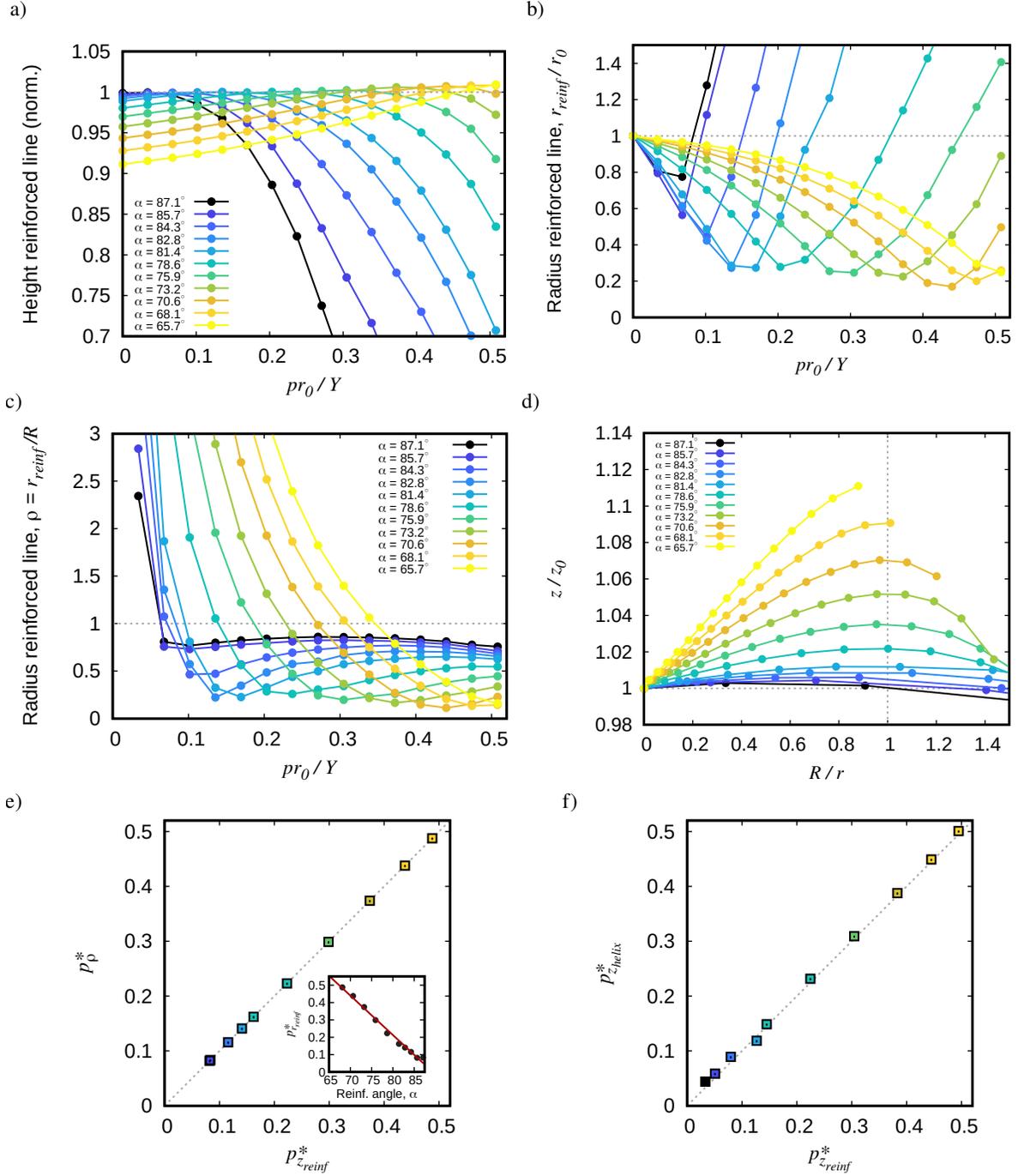

FIG. S19. Configuration of the reinforced line during pressurization. a) Maximal height of the reinforced string normalized by the theoretical contour length of the string $L_f = L_{c_0}\sqrt{1+\cot^2\alpha}$. Detail showing the initial linear increase of the height with pressure as a result of straightening with the long axis before shortening. b) Average radius of the reinforced line $r_{reinf}$, measured with respect to the long axis, normalized by the initial tube radius $r_0$. c) Same as in panel b) but normalized by the radius of the helix, where we define $\rho = r_{reinf}/R$. Values of $\rho < 1$ indicate a localization of the reinforced string on the internal region of the helix. d) Height as a function of the ratio between the radius of the tube $r$ and the radius of the helix $R$. The maximum height is reached for $R \approx r$. e) Critical pressure $p_\rho^*$ of occurrence of the minimal radius of the reinforced string as a function of the pressure of maximum string height, $p_{z_{reinf}}^*$. *Inset.* Dependency of $p_{r_{reinf}}^*$ on the reinforcement angle. Red line is a linear fit. f) Pressure of maximal height of the helix main body $p_z^*$ as a function of $p_{z_{reinf}}^*$. The dashed lines with slope one in e) and f) are a guide to the eye.

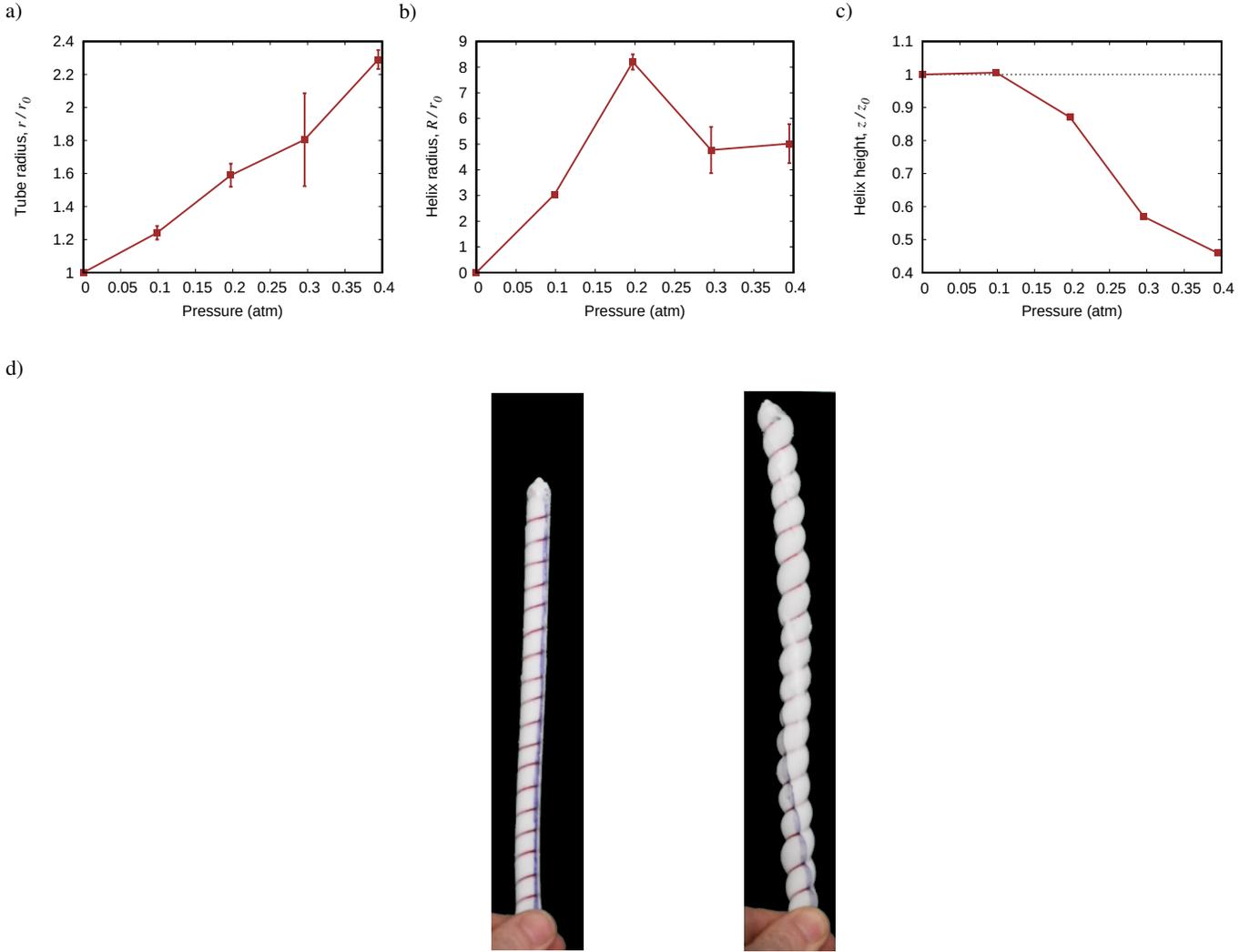

FIG. S20. Helical properties of reinforced balloons. a) Tube radius $r$, b) helix radius $R$, and c) helical height $z$ as a function of the pressure of pressurized balloons with steep reinforcement angle ($\alpha = 73°$). Error bars are the standard deviation of measurements on different regions of the pressurized balloon shown in the main text. d) Low reinforcing angles ($\alpha = 15°$) show clearly noticeable torsion but no height reduction. *Left.* Initial configuration at zero pressure. *Right.* Configuration at $p \approx 2$ atm. The reinforced string is visible in red. The line in blue is a guide to the eye, drawn as a straight line in the undeformed configuration.


\fi

\end{document}